\documentclass[11pt]{iopart}
 \pdfoutput=1
 \usepackage{tikz}
 \usepackage{amsthm,graphicx,amsfonts,amssymb,latexsym,cancel,verbatim,color,amscd}
  \eqnobysec

 \theoremstyle{plain}
 \newtheorem {hypo}{\bf\hspace{-\parindent}Hypothesis}

 \newtheorem {conj}[hypo]{Conjecture}
 \theoremstyle{definition}
 \theoremstyle{remark}
 \newtheorem {rmk}[hypo]{Remark}%[section]

 \newcommand\ben{\begin{equation*}}
 \newcommand\ebn{\end{equation*}}
 \newcommand\be{\begin{equation}}
 \newcommand\eb{\end{equation}}

 \newcommand{\pf}{\begin{bpf}}

 \newcommand{\pfms}{\begin{bpfms}}
 \newcommand{\epf}{\end{bpf}\hfill$\square$\vspace{0.1cm}}
 \newcommand{\epfms}{\end{bpfms}\hfill$\square$\\ }

\begin{document}
 \title{How instanton combinatorics solves Painlev\'e~VI,~V~and~III's}
 \author{O. Gamayun$^{1,2}$, N. Iorgov$^{1}$, O. Lisovyy$^{1,3}$}

% The "\note" macro will give a warning: "Ignoring empty anchor..."
% you can safely ignore it.

\address{$^1$ Bogolyubov Institute for Theoretical Physics,
03680, Kyiv, Ukraine}
\address{$^2$ Physics Department, Lancaster University,
Lancaster, LA1 4YB, United Kingdom}
\address{$^3$ Laboratoire de Math\'ematiques et Physique Th\'eorique CNRS/UMR 7350,
 Universit\'e de Tours,
 37200 Tours, France}
 \eads{\mailto{o.gamayun@lancaster.ac.uk}, \mailto{iorgov@bitp.kiev.ua}, \mailto{lisovyi@lmpt.univ-tours.fr}}

%  \maketitle

 \begin{abstract}
 We elaborate on a recently conjectured relation of Painlev\'e transcendents and 2D CFT.
 General solutions of Painlev\'e~VI, V and III are expressed in terms of  $c=1$ conformal
 blocks  and their irregular limits, AGT-related to instanton partition functions in
 $\mathcal{N}=2$ supersymmetric gauge theories with $N_f=0,1,2,3,4$. Resulting combinatorial series representations of
 Painlev\'e functions provide
 an efficient tool for their numerical computation at finite
 values of the argument. The series involve sums over bipartitions which in the simplest cases
 coincide with Gessel expansions of certain Toeplitz determinants.
 Considered applications include Fredholm determinants of classical integrable kernels, scaled gap probability in the bulk of the GUE,
 and all-order conformal perturbation theory expansions
 of correlation functions in the sine-Gordon field theory at the free-fermion point.
 \end{abstract}
  \vspace{0.5cm}

 \date{}

 \section{Introduction}
 Painlev\'e transcendents \cite{clarkson} are nowadays widely recognized as important special functions with a broad range
 of applications including integrable models, combinatorics and random matrix theory. Many aspects of Painlev\'e equations,
 such as their analytic and geometric properties, asymptotic problems, special solutions and discretization,
 have been extensively studied in the last four decades.

 From the point of view of the theory of classical special functions \cite{VK},
 the surprising feature of these developments  is the absence of transparent connection to representation theory.
 Instead, the Riemann-Hilbert approach \cite{fokas} is typically used.
 It is well-known that Painlev\'e equations emerge most naturally in the study of monodromy preserving deformations of
  linear ODEs. Thus, by analogy with the solution of classical integrable systems
 by the inverse scattering method, the questions on nonlinear Painlev\'e functions may  be asked
  in terms of linear monodromy. In particular,
 one may attempt to realize the following program:
 \begin{itemize}
 \item label different Painlev\'e functions by monodromy data of the auxiliary linear problem,
 \item express their asymptotics near the critical points in terms of monodromy,
 \item construct full solution using the asymptotic behaviour as initial condition.
 \end{itemize}

 Starting from the foundational work of Jimbo \cite{jimbo}, there are many results available on the first two points,
 but the lack of algebraic structure makes the last one
 difficult to tackle. In other words, the question is
 \begin{itemize}
 \item[] \textit{...how does one combine asymptotic information about the solutions obtained
 from the Riemann-Hilbert problem, together with efficient numerical codes in order
 to compute the solution $u(x)$ at finite values of $x$?} \cite[Painlev\'e Project Problem]{deift}.
 \end{itemize}

 In \cite{CFT_PVI}, a solution of this problem was suggested for the sixth Painlev\'e equation. It was shown that
 Painlev\'e~VI tau function $\tau_{_{\mathrm{VI}}}(t)$ can be thought of as a correlation function of primary fields in
 2D conformal field theory  \cite{BPZ} with central charge $c=1$. Under natural minimal assumptions on
 primary content of the theory and fusion rules, $\tau_{_{\mathrm{VI}}}(t)$ may then
 be written as a linear combination of Virasoro conformal blocks.
 Being purely representation-theoretic quantities, these CFT special functions can be computed in several ways.
 In particular, the recently proven \cite{AGT_proof} AGT conjecture \cite{AGT} relates them to instanton partition
 functions in $\mathcal{N}=2$ SUSY 4D Yang-Mills theories \cite{Bruzzo,FlP,Nekrasov1,Nekrasov_Okounkov},
 which can be expressed as sums over tuples of partitions. This results into combinatorial series representations
 of $\tau_{_{\mathrm{VI}}}(t)$ around the critical points $0$, $1$, $\infty$.

 The aim of this note is to extend the results of \cite{CFT_PVI} to Painlev\'e~V and Painlev\'e~III, and to make them
 accessible to a wider audience. With this purpose in mind, we deliberately include some background material and
 illustrate our claims with a number of explicit examples
 and applications to random matrix theory and integrable QFT.

 The plan is as follows. Section~\ref{PE} sets the notation and explains the relation between different
 Painlev\'e equations and their various forms. In Section~\ref{seccft}, we recall some basics on conformal blocks and
 AGT correspondence. Conjectural general solutions of Painlev\'e~VI, V and III are presented and discussed in Section~\ref{secsols}.
 In particular, it is shown that our combinatorial expansions can be seen as a generalization
 of Gessel's theorem representation of classical Toeplitz determinant solutions.
 Section~\ref{secapp} is devoted to applications, which include Fredholm determinants of classical integrable kernels
 (hypergeometric, Whittaker, confluent hypergeometric and modified Bessel), scaled GUE bulk gap probability
  and correlators of exponential fields
 in the sine-Gordon model at the free-fermion point.

 \section{Painlev\'e equations}\label{PE}
 \subsection{Conventional form}
 Painlev\'e VI, V, and III ($P_{_{\mathrm{VI}}}$, $P_{_{\mathrm{V}}}$, $P_{_{\mathrm{III}}}$) first appeared as
 a part of the classification of 2nd order, 1st degree nonlinear
 ODEs without movable critical points. In this context, they are usually written as follows:\vspace{0.1cm}\\
 \textit{Painlev\'e VI}:
 \begin{eqnarray}\label{pvis}
 \fl\qquad\frac{d^2q}{dt^2}=\frac{1}{2}\left(\frac{1}{q}+\frac{1}{q-1}
 +\frac{1}{q-t}\right)\left(\frac{dq}{dt}\right)^2 -
 \left(\frac{1}{t}+\frac{1}{t-1}+\frac{1}{q-t}\right)\frac{dq}{dt}\,+\vspace{0.1cm}\\
 \nonumber  +\,
 \frac{2q(q-1)(q-t)}{t^2(t-1)^2}\left(\alpha+\frac{\beta t}{q^2}+
 \frac{\gamma(t-1)}{(q-1)^2}+\frac{\delta t(t-1)}{(q-t)^2}\right),
 \end{eqnarray}
 \textit{Painlev\'e V}:
 \begin{eqnarray}\label{pvs}
 \fl\qquad\frac{d^2q}{dt^2}=\left(\frac{1}{2q}+\frac{1}{q-1}\right)\left(\frac{dq}{dt}\right)^2 -
 \frac{1}{t}\frac{dq}{dt}\,+\frac{(q-1)^2}{t^2}\left(\alpha q+\frac{\beta}{q}\right)+\frac{\gamma q}{t}
 +\frac{\delta q(q+1)}{q-1},
 \end{eqnarray}
  \textit{Painlev\'e III}:
 \begin{eqnarray}\label{piiis}
 \fl\qquad\frac{d^2q}{dt^2}=\frac{1}{q}\left(\frac{dq}{dt}\right)^2 -
 \frac{1}{t}\frac{dq}{dt}\,+\frac{\alpha q^2+\beta}{t}+\gamma q^3+\frac{\delta}{q}.
 \end{eqnarray}
  It is often
  convenient to use instead of $P_{_{\mathrm{III}}}$ an equivalent equation,\vspace{0.1cm}\\
 \textit{Painlev\'e III$\,'$}:
   \begin{eqnarray}\label{piiispr}
 \fl\qquad\frac{d^2q}{dt^2}=\frac{1}{q}\left(\frac{dq}{dt}\right)^2 -
 \frac{1}{t}\frac{dq}{dt}\,+\frac{q^2\left(\alpha+\gamma q\right)}{4t^2}+\frac{\beta}{4t}+\frac{\delta}{4q},
 \end{eqnarray}
 which reduces to $P_{_{\mathrm{III}}}$ by setting $t_{_{\mathrm{III'}}}=t_{_{\mathrm{III}}}^2$,
 $q_{_{\mathrm{III'}}}=t_{_{\mathrm{III}}}q_{_{\mathrm{III}}}$.

 \subsection{Parameterization}
   We write four $P_{_{\mathrm{VI}}}$ parameters as
 \begin{eqnarray}
 \label{parsp6}
 \left(\alpha,\beta,\gamma,\delta\right)_{_{\mathrm{VI}}}=\left(\Bigl(\theta_{\infty}+\frac12\Bigr)^2,-\theta_0^2,\theta_1^2,
 \frac14-\theta_t^2\right).
 \end{eqnarray}

 If $\delta\neq0$ in $P_{_{\mathrm{V}}}$, then one can set $\delta=-\frac12$ by rescaling the independent variable.
 $P_{_{\mathrm{V}}}$ with $\delta=0$ is reducible to $P_{_{\mathrm{III}}}$ (see e.g. transformations (1.24)--(1.26) in \cite{cosgrove})
 which will be treated separately. Hence we may set
 \begin{eqnarray}
 \label{parsp5}
 \left(\alpha,\beta,\gamma,\delta\right)_{_{\mathrm{V}}}=\Bigl(2\theta_0^2,-2\theta_t^2,2\theta_*-1,
 -\frac12\Bigr).
 \end{eqnarray}

 The case of $P_{_{\mathrm{III}}}$ is slightly more involved. In the generic situation, when $\gamma\delta\neq0$, one can
 assume that $\gamma=-\delta=4$ by rescaling $t$ and $q$. We will then write
  \begin{eqnarray}\label{parsp31}
 \left(\alpha,\beta,\gamma,\delta\right)_{_{\mathrm{III}_1}}=\left(8\theta_{\star},4-8\theta_*,4,
 -4\right).
 \end{eqnarray}
 The variable change $q\rightarrow q^{-1}$ maps $P_{_{\mathrm{III}}}$ with $\delta=0$ to $P_{_{\mathrm{III}}}$ with $\gamma=0$.
 Assume that $\gamma=0$ and $\alpha\delta\neq0$, then the scaling freedom can be used to set
   \begin{eqnarray}\label{parsp32}
 \left(\alpha,\beta,\gamma,\delta\right)_{_{\mathrm{III}_2}}=\left(8,4-8\theta_*,0,
 -4\right).
 \end{eqnarray}
 For $\gamma=\delta=0$, $\alpha\beta\neq 0$ we can set
 \begin{eqnarray}\label{parsp33}
 \left(\alpha,\beta,\gamma,\delta\right)_{_{\mathrm{III}_3}}=\left(8,-8,0, 0\right).
 \end{eqnarray}
 Finally, for $\alpha=\gamma=0$ (and, similarly, for $\beta=\delta=0$ by $q\rightarrow q^{-1}$), the general (two-parameter) solution of $P_{_{\mathrm{III}}}$ is known \cite{lukash}. It reads
 \ben
 q(t)=\mu t^{1-\nu}+\frac{\beta}{\nu^{2}}\,t+\frac{\beta^2+\nu^2\delta}{4\mu\nu^4}\,t^{1+\nu},
 \ebn
 where $\mu$, $\nu$ are two arbitrary integration constants. Excluding this last solvable case,
 there remain three inequivalent $P_{_{\mathrm{III}}}$'s with two, one and zero parameters.
 Significance of the degenerate equations $P_{_{\mathrm{III}_2}}$ and $P_{_{\mathrm{III}_3}}$ was realized in \cite{sakai} from a geometric viewpoint, and later they were extensively studied in \cite{ohyama}.

 \subsection{Hamiltonian form}
 Painlev\'e equations can be written as non-autonomous hamiltonian systems \cite{malm}. In this
 approach, (\ref{pvis}), (\ref{pvs}) and (\ref{piiispr}) are obtained by eliminating momentum $p$ from the equations
 \ben
 \frac{dq}{dt}=\frac{\partial H_{_{\mathrm{J}}}}{\partial p},\qquad \frac{dp}{dt}=-\frac{\partial H_{_{\mathrm{J}}}}{\partial q},\qquad\qquad \mathrm{J}=\mathrm{VI},\mathrm{V},\mathrm{III}'_{1,2,3},
 \ebn
 where the  Hamiltonians are given by
 \begin{eqnarray}
 \nonumber t(t-1)H_{_{\mathrm{VI}}}=q\left(q-1\right)\left(q-t\right)p\left(p-\frac{2\theta_0}{q}-\frac{2\theta_1}{q-1}-
 \frac{2\theta_t-1}{q-t}\right)+\\
 \label{hamp6}\qquad\qquad +\left(\theta_0+\theta_t+\theta_1+\theta_{\infty}\right)
 \left(\theta_0+\theta_t+\theta_1-\theta_{\infty}-1\right)q,\\
 \nonumber tH_{_{\mathrm{V}}}=(q-1)(pq-2\theta_t)(pq-p+2\theta_*)-tpq+
 \label{hamp5}\left(\left(\theta_*+\theta_t\right)^2-\theta_0^2\right)q+\\
 \qquad\qquad +\Bigl(\theta_t-\frac{\theta_*}{2}\Bigr)t-
 2\Bigl(\theta_t+\frac{\theta_*}{2}\Bigr)^2,\\
 \label{hamp31} tH_{_{\mathrm{III}_1'}}=\left(pq+\theta_{*}\right)^2+tp-\theta_{\star}q-\frac{q^2}{4},\\
 \label{hamp32} tH_{_{\mathrm{III}_2'}}=\left(pq+\theta_{*}\right)^2+tp-q,\\
 \label{hamp33} tH_{_{\mathrm{III}_3'}}=p^2q^2-q-\frac{t}{q}.
 \end{eqnarray}
 The hamiltonian structure is crucial for the construction of
 Okamoto-B\"acklund trans\-for\-mations \cite{okamoto86}, generating
 an infinite number of Painlev\'e solutions from a given one.

 \subsection{Sigma form and tau functions}
 The time-dependent Hamiltonians (\ref{hamp6})--(\ref{hamp33}) themselves satisfy  nonlinear 2nd order ODEs. To write them,
 introduce auxiliary functions
 \begin{eqnarray}
 \label{sigmap6}
 \nonumber\sigma_{_{\mathrm{VI}}}=
 t(t-1)H_{_{\mathrm{VI}}}-q(q-1)p+\left(\theta_0+\theta_t+\theta_1+\theta_{\infty}\right)q\\
 \qquad -\left(\theta_0+\theta_1\right)^2 t+\frac{\theta_1^2+\theta_{\infty}^2-\theta_0^2-\theta_t^2-4\theta_0\theta_t}{2},\\
  \sigma_{_{\mathrm{J}}}=tH_{_{\mathrm{J}}},\qquad\qquad\qquad \mathrm{J=V,III'_{1,2,3}}.
 \end{eqnarray}
 They satisfy the so-called $\sigma$-form of Painlev\'e equations \cite{Forrester_book,jm2}:
 \begin{eqnarray}
% \tag{$\sigma$PVI}
 \label{sigpvi}\fl P_{_{\mathrm{VI}}}:\qquad\sigma'\Bigl(t(t-1)\sigma''\Bigr)^2+\left[2\sigma'(t\sigma'-\sigma)-\left(\sigma'\right)^2-
 (\theta_t^2-\theta_{\infty}^2)(\theta_0^2-\theta_1^2)\right]^2= \\
 \nonumber\fl\qquad\qquad = \left(\sigma'+\left(\theta_t+\theta_{\infty}\right)^2\right)
 \left(\sigma'+\left(\theta_t-\theta_{\infty}\right)^2\right)
 \left(\sigma'+\left(\theta_0+\theta_{1}\right)^2\right)
 \left(\sigma'+\left(\theta_0-\theta_{1}\right)^2\right),\\
 %\tag{$\sigma$PV}
 \label{sigpv}\fl P_{_{\mathrm{V}}}:\qquad\left(t\sigma''\right)^2=\left(\sigma-t\sigma'+2\left(\sigma'\right)^2\right)^2\!\! -
 \frac14\left(\left(2\sigma'-\theta_*\right)^2\!\!-4\theta_0^2\right)
 \left(\left(2\sigma'+\theta_*\right)^2\!\!-4\theta_t^2\right),\\
% \tag{$\sigma$PIII}
 \label{sigpiii1}\fl P_{_{\mathrm{III_1'}}}:\quad\;\;
 \left(t\sigma''\right)^2=\bigl(4\left(\sigma'\right)^2-1\bigr)\left(\sigma-t\sigma'\right)-4\theta_*\theta_{\star}\sigma'+
 \left(\theta_*^2+\theta_{\star}^2\right),\\
% \tag{$\sigma$PIII$'$}
 \label{sigpiii2}\fl P_{_{\mathrm{III_2'}}}:\quad\;\;
 \left(t\sigma''\right)^2=4\left(\sigma'\right)^2\left(\sigma-t\sigma'\right)-4\theta_*\sigma'+1,\\
 %\tag{$\sigma$PIII$''$}
 \label{sigpiii3}\fl P_{_{\mathrm{III_3'}}}:\quad\;\;
 \left(t\sigma''\right)^2=4\left(\sigma'\right)^2\left(\sigma-t\sigma'\right)-4\sigma',
 \end{eqnarray}
 which also appear in the classification of 2nd order, 2nd degree ODEs with Painlev\'e property \cite{cosgrove}.

 The solutions of (\ref{pvis})--(\ref{piiispr})
 can thus be mapped to solutions of (\ref{sigpvi})--(\ref{sigpiii3}).
 Conversely, one can recover conventional Painlev\'e functions from the
 solutions of $\sigma$-Painlev\'e equations using the following formulas:
 \begin{eqnarray}
 \label{invpvi}\fl P_{_{\mathrm{VI}}}:\qquad\; \frac{1}{q-t}+\frac12\left(\frac1t+\frac{1}{t-1}\right)=\\
 \nonumber\fl\qquad\qquad=
 \frac{2\theta_{\infty}t(t-1)\sigma''+\left(\sigma'+\theta_t^2-\theta_{\infty}^2\right)\left(\left(2t-1\right)\sigma'-2\sigma
 +\theta_0^2-\theta_1^2\right)+4\theta_{\infty}^2\left(\theta_0^2-\theta_1^2\right)
 }{2t(t-1)\left(\sigma'+\left(\theta_{t}-\theta_{\infty}\right)^2\right)
 \left(\sigma'+\left(\theta_{t}+\theta_{\infty}\right)^2\right)},\\
 \label{invpv}\fl P_{_{\mathrm{V}}}:\qquad \;\; q=\frac{2\left(t\sigma''+\sigma-t\sigma'+2\left(\sigma'\right)^2\right)}{\left(2\sigma'-\theta_*\right)^2-4\theta_0^2},\\
 \label{invpiii1}\fl P_{_{\mathrm{III}'_1}}:\qquad q=-\frac{2t\sigma''+4\theta_*\sigma'-2\theta_{\star}}{4\left(\sigma'\right)^2-1},\\
 \label{invpiii2}\fl P_{_{\mathrm{III}'_2}}:\qquad q=-\frac{t\sigma''+2\theta_*\sigma'-1}{2\left(\sigma'\right)^2},\\
 \label{invpiii3}\fl P_{_{\mathrm{III}'_3}}:\qquad q=-\frac{1}{\sigma'}.
 \end{eqnarray}

 Finally, define the tau functions of $P_{_{\mathrm{VI}}}$, $P_{_{\mathrm{V}}}$ and $P_{_{\mathrm{III}}}$ by
 \begin{eqnarray}
 \label{tauvi}
 \sigma_{_{\mathrm{VI}}}(t)=&\;t(t-1)\frac{d}{dt}\ln \left( t^{\frac{\theta_0^2+\theta_t^2-\theta_1^2-\theta^2_{\infty}}{2}}
 \left(1-t\right)^{\frac{\theta^2_{t}+\theta_1^2-\theta_0^2-\theta_{\infty}^2}{2}}\tau_{_{\mathrm{VI}}}(t)\right),\\
 \label{tauv}
 \sigma_{_{\mathrm{V}}}(t)=&\;t\frac{d}{dt}\ln\left(e^{-\frac{\theta_* t}{2}}t^{-\theta_0^2-\theta_t^2-\frac{\theta_*^2}{2}}\tau_{_{\mathrm{V}}}(t)\right),\\
 \label{tauiii}
 \sigma_{_{\mathrm{J}}}(t)=&\;t\frac{d}{dt}\ln\tau_{_{\mathrm{J}}}(t),\qquad\qquad \mathrm{J=III'_{1,2,3}}.
 \end{eqnarray}
 Our solution below is formulated in terms of combinatorial expansions of these
 tau functions in powers of $t$. Expansions of $\sigma$'s and $q$'s can then be
 obtained from the relations (\ref{invpvi})--(\ref{invpiii3}) and (\ref{tauvi})--(\ref{tauiii}).
 \subsection{Coalescence}
 As is well-known, Painlev\'e~VI produces all other Painlev\'e equations in certain scaling limits.
 The equations considered in the present paper form the first line of the coalescence cascade

 \begin{center}
 \begin{tikzpicture}[node distance=1.8cm, auto]
 \node (P6) {$\mathrm{VI}$};
 \node (P5) [right of=P6] {$\mathrm{V}$};
 \node (P31) [right of=P5] {$\mathrm{III_1}$};
 \node (P32) [right of=P31] {$\mathrm{III_2}$};
 \node (P33) [right of=P32] {$\mathrm{III_3}$};
 \node (P4) [below of=P5] {$\mathrm{IV}$};
 \node (P2) [below of=P31] {$\mathrm{II}$};
 \node (P1) [below of=P32] {$\mathrm{I}$};
 \draw[->] (P6) to node {} (P5);
  \draw[->] (P5) to node {} (P31);
   \draw[->] (P31) to node {} (P32);
    \draw[->] (P32) to node {} (P33);
     \draw[->] (P5) to node {} (P4);
      \draw[->] (P4) to node {} (P2);
       \draw[->] (P2) to node {} (P1);
        \draw[->] (P31) to node {} (P2);
         \draw[->] (P32) to node {} (P1);
 \end{tikzpicture}\\
 Fig. 1: Coalescence diagram for Painlev\'e equations
 \end{center}
 Every step to the right or to the bottom of the diagram decreases by 1 the number of parameters in
 the corresponding equation.

 Let us now describe the scaling limits we need (1st line) and the transition $P_{_{\mathrm{III}_2}}\rightarrow P_{_{\mathrm{I}}}$
 which seems to be missing in the literature (cf. e.g. the degeneration scheme in \cite{ohyama}).
 \begin{itemize}
 \item $P_{_{\mathrm{VI}}}\rightarrow P_{_{\mathrm{V}}}$: set in $P_{_{\mathrm{VI}}}$
 \begin{eqnarray}
 \label{limpvitopv1}\theta_1=\frac{\Lambda+\theta_*}{2},\qquad
 \theta_{\infty}=\frac{\Lambda-\theta_*}{2},
 \end{eqnarray}
 then solutions of $P_{_{\mathrm{V}}}$ can be obtained as the limits
 \begin{eqnarray}
 \label{limpvitopv2}1-q_{_{\mathrm{V}}}\left(t\right)=
 \lim_{\Lambda\rightarrow\infty}\frac{t/\Lambda}{q_{_{\mathrm{VI}}}\left(t/\Lambda\right)},\\
 \label{limpvitopv3}
 \sigma_{_{\mathrm{V}}}\left(t\right)=\lim_{\Lambda\rightarrow\infty}
 \left(\frac{\Lambda^2-\Lambda t-\theta_*^2-2\theta_0^2-2\theta_t^2}{4}
 -\sigma_{_{\mathrm{VI}}}\left(t/\Lambda\right)\right),\\
  \label{limpvitopv4}\tau_{_{\mathrm{V}}}\left(t\right)=
 \lim_{\Lambda\rightarrow\infty}\left({t}/{\Lambda}\right)^{\theta_0^2+\theta_t^2}\tau_{_{\mathrm{VI}}}\left({t}/{\Lambda}\right),
 \end{eqnarray}
 \item $P_{_{\mathrm{V}}}\rightarrow P_{_{\mathrm{III}'_1}}$: this limiting transition is described by
 \begin{eqnarray}
 \label{pvtopiii1first}
 \theta_{0}=\frac{\Lambda-\theta_{\star}}{2},\qquad \theta_t=\frac{\Lambda+\theta_{\star}}{2},\\
 q_{_{\mathrm{III}'_1}}\left(t\right)\,=
 \lim_{\Lambda\rightarrow\infty}\Lambda\left(1-q_{_{\mathrm{V}}}\left({t}/{\Lambda}\right)\right),\\
 \sigma_{_{\mathrm{III}'_1}}\left(t\right)=
 \lim_{\Lambda\rightarrow\infty}\left(\frac{\Lambda^2+\theta_*^2+\theta_{\star}^2}{2}+\sigma_{_{\mathrm{V}}}\left(t/\Lambda\right)\right),\\
 \label{pvtopiii1last}
 \tau_{_{\mathrm{III}'_1}}\left(t\right)\,=
 \lim_{\Lambda\rightarrow\infty}\tau_{_{\mathrm{V}}}\left({t}/{\Lambda}\right),
 \end{eqnarray}
 \item $P_{_{\mathrm{III}'_1}}\rightarrow
 P_{_{\mathrm{III}'_2}}$: similarly,
 \begin{eqnarray}
 \label{piii1topiii2first}
 q_{_{\mathrm{III}'_2}}\left(t\right)\,=
 \lim_{\theta_{\star}\rightarrow\infty}\theta_{\star}q_{_{\mathrm{III}'_1}}\left({t}/{\theta_{\star}}\right),\\
 \sigma_{_{\mathrm{III}'_2}}\left(t\right)=
 \lim_{\theta_{\star}\rightarrow\infty}\sigma_{_{\mathrm{III}'_1}}\left({t}/{\theta_{\star}}\right),\\
 \label{piii1topiii2last}
 \tau_{_{\mathrm{III}'_2}}\left(t\right)\,=
 \lim_{\theta_{\star}\rightarrow\infty}\tau_{_{\mathrm{III}'_1}}\left({t}/{\theta_{\star}}\right).
 \end{eqnarray}
 \item $P_{_{\mathrm{III}'_2}}\rightarrow
 P_{_{\mathrm{III}'_3}}$:
  \begin{eqnarray}
  \label{piii2topiii3first}
 q_{_{\mathrm{III}'_3}}\left(t\right)\,=
 \lim_{\theta_{*}\rightarrow\infty}q_{_{\mathrm{III}'_2}}\left({t}/{\theta_{*}}\right),\\
 \sigma_{_{\mathrm{III}'_3}}\left(t\right)=
 \lim_{\theta_{*}\rightarrow\infty}\sigma_{_{\mathrm{III}'_2}}\left({t}/{\theta_{*}}\right),\\
  \label{piii2topiii3last}
 \tau_{_{\mathrm{III}'_3}}\left(t\right)\,=
 \lim_{\theta_{*}\rightarrow\infty}\tau_{_{\mathrm{III}'_2}}\left({t}/{\theta_{*}}\right).
 \end{eqnarray}
 \item $P_{_{\mathrm{III}'_2}}\rightarrow P_{_{\mathrm{I}}}$: set
 \begin{eqnarray}
 \label{transp32top1e1}{\theta_*}=3\Lambda^{\frac54},\qquad t_{_{\mathrm{III}'_2}}=16\Lambda^{\frac{15}{4}}\left(1+\frac{t}{2\Lambda}\right),\\
 \label{transp32top1e2}\sigma_{_{\mathrm{III}'_2}}\left(t_{_{\mathrm{III}'_2}}\right)=2\Lambda\sigma(t)
 +8\Lambda^{\frac52}+\frac{1}{4}\Lambda^{-\frac54} t_{_{\mathrm{III}'_2}},
 \end{eqnarray}
 then in the limit $\Lambda\rightarrow\infty$ the function $\sigma(t)$ satisfies the $\sigma$-form of $P_{_{\mathrm{I}}}$, namely,
 \be\label{spi}
 \left(\sigma''\right)^2=2\sigma-2t\sigma'-4\left(\sigma'\right)^3.
 \eb
 Also, if (\ref{transp32top1e2}) is replaced with
 \be\label{transp32top1e3}
 q_{_{\mathrm{III}'_2}}\left(t_{_{\mathrm{III}'_2}}\right)=-4\Lambda^{\frac52}+4\Lambda^2 q(t),
 \eb
 the limiting equation for $q(t)$ is $P_{_{\mathrm{I}}}$ in the conventional form:
 \be\label{convpi}
 q''=6q^2+t.
 \eb
 \end{itemize}

 \subsection{Analytic properties}
 The only branch points of $P_{_{\mathrm{VI}}}$ and $P_{_{\mathrm{V,III'_{1,2,3}}}}$ transcendents in the extended complex
 $t$-plane  are $0,1,\infty$ and $0,\infty$, res\-pec\-tively. The corresponding tau functions are holomorphic on the universal covers
 of $\mathbb{P}^1\backslash\{0,1,\infty\}$ and $\mathbb{P}^1\backslash\{0,\infty\}$.
 The functions $q$ and $\sigma$ may also have movable poles associated to zeros of $\tau$. We introduce the branch cuts
 $(-\infty,0]\cup[1,\infty)$ (for $P_{_{\mathrm{VI}}}$) and $(-\infty,0]$ (for $P_{_{\mathrm{V,III'_{1,2,3}}}}$) and adopt the
 principal branch convention for all fractional powers of $t$ and $1-t$.

 \section{Conformal blocks and instanton partition functions}\label{seccft}
 \subsection{Conformal blocks}
 Here we review basic notions about conformal blocks in 2D CFT \cite{BPZ}. For the sake of brevity,
 simplicity and relevance for the rest of the presentation, we will concentrate on conformal blocks for the 4-point correlator
 on $\mathbb{P}^1$ and Virasoro algebra
 \ben
 [L_n,L_m]=\left(n-m\right)L_{n+m}+\frac{c}{12}\left(n^3-n\right)\delta_{n+m,0}.
 \ebn
 Only chiral primary fields $\mathcal{O}$ will be considered, i.e. their
 antiholomorphic conformal dimensions $\bar{\Delta}_{\mathcal{O}}=0$. We do not require invariance of correlators under the
 braid group action on the positions of fields to avoid constraints on dimensions.

 The three-point correlator of primary fields is fixed by conformal symmetry up to a constant factor,
 \ben
 \langle\mathcal{O}_3(z_3)\mathcal{O}_2(z_2)\mathcal{O}_1(z_1)\rangle=C\left(\Delta_3,\Delta_2,\Delta_1\right)
 z_{21}^{\Delta_3-\Delta_1-\Delta_2}z_{32}^{\Delta_1-\Delta_2-\Delta_3}z_{31}^{\Delta_2-\Delta_1-\Delta_3},
 \ebn
 where $z_{ij}=z_i-z_j$ and $\Delta_j$ stand for holomorphic dimensions. Thanks to conformal invariance, it suffices to
 consider more special coordinate dependence, namely, we may set $z_1=0$, $z_2= t$, $z_3=R$ with $R\gg t$.
% Then, to leading order in $R$,
% \ben
% \langle\mathcal{O}_3(R)\mathcal{O}_2(t)\mathcal{O}_1(0)\rangle\simeq
% C\left(\Delta_3,\Delta_2,\Delta_1\right)R^{-2\Delta_3}t^{\Delta_3-\Delta_1-\Delta_2}.
% \ebn
 It is also customary to define $\langle\mathcal{O}(\infty)\ldots\rangle=\lim_{R\rightarrow\infty}R^{2\Delta_{\mathcal{O}}}\langle\mathcal{O}(R)\ldots\rangle$, so that,
 for instance, $\langle\mathcal{O}_3(\infty)\mathcal{O}_2(t)\mathcal{O}_1(0)\rangle=C\left(\Delta_3,\Delta_2,\Delta_1\right)
 t^{\Delta_3-\Delta_1-\Delta_2}$.

 Besides primary fields, conformal field theory also contains their descendants
 $L_{-\lambda}\mathcal{O}=L_{-\lambda_N}\ldots L_{-\lambda_1}\mathcal{O}$, naturally labeled by partitions $\lambda=\left\{\lambda_1\geq\lambda_2\geq\ldots\geq\lambda_N>0\right\}$.
 Partitions can be identified in the obvious way with Young diagrams. As they play an important role in the rest of the
 paper, we take the opportunity to fix some notation for later purposes. The set of all Young diagrams will
 be denoted by $\mathbb{Y}$. For $\lambda\in\mathbb{Y}$, $\lambda'$ denotes the transposed diagram, $\lambda_i$
 and $\lambda'_j$ the number of boxes in $i$th row and $j$th column of $\lambda$, and $|\lambda|$
 the total number of boxes. Given a box $(i,j)\in\lambda$, its hook length is defined as $h_{\lambda}(i,j)=
 \lambda_i+\lambda'_j-i-j+1$
  (see Fig.~2).

    \begin{figure}[!h]
 \begin{center}
 \resizebox{7cm}{!}{
 \includegraphics{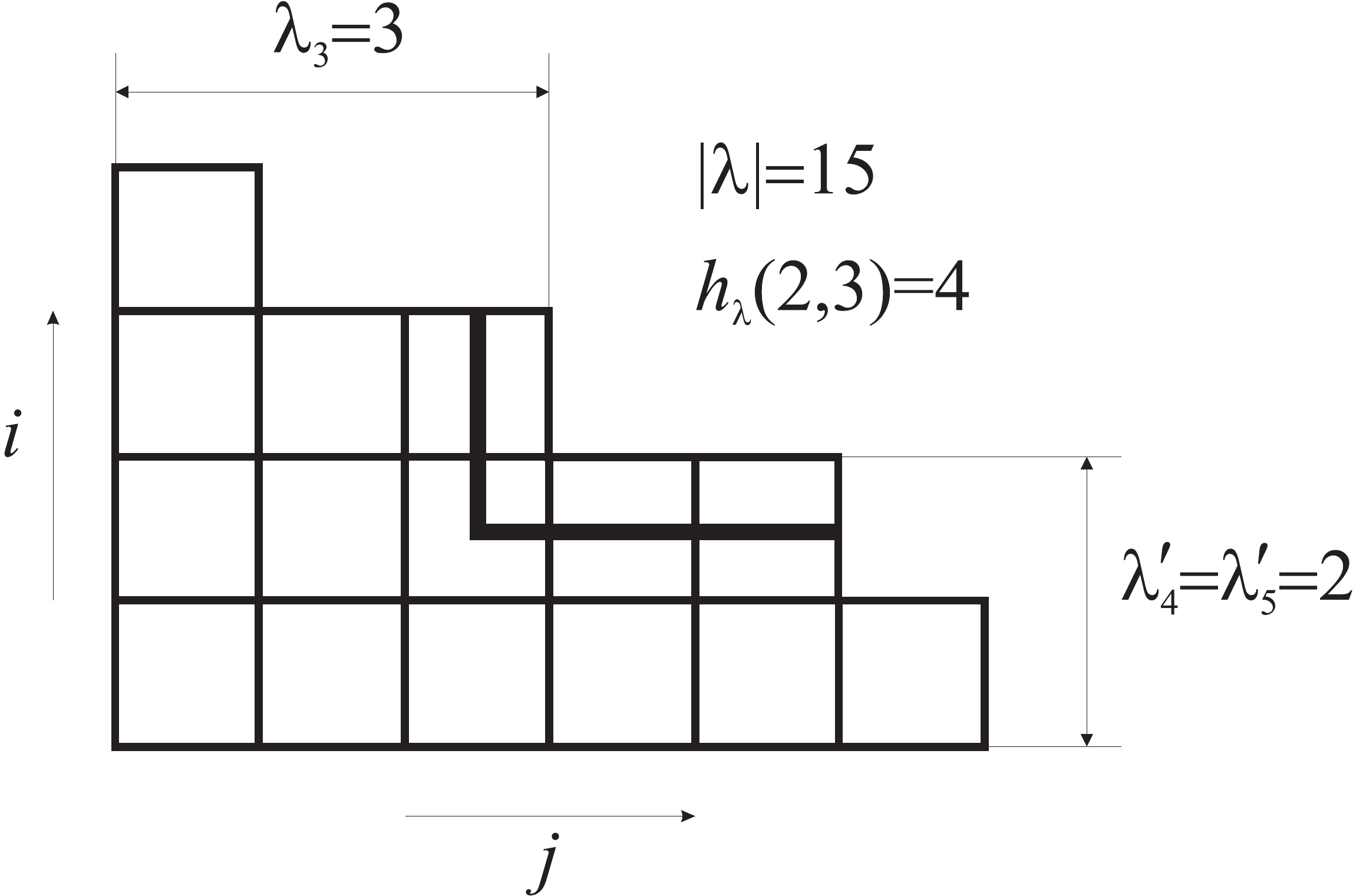}} \\
 Fig. 2: Young diagram associated to the partition $\lambda=\{6,5,3,1\}$
 \end{center}
 \end{figure}

 Conformal symmetry also allows to compute in explicit form
 the three-point functions involving one descendant:
 \begin{eqnarray}
 \label{imprel0}
 \langle \mathcal{O}_3\left(\infty\right)\mathcal{O}_2(t)L_{-\lambda}\mathcal{O}_1\left(0\right)\rangle=
 C\left(\Delta_3,\Delta_2,\Delta_1\right)
 \gamma_{\lambda}\left(\Delta_1,\Delta_2,\Delta_3\right)t^{-|\lambda|},\\
 \label{imprel}
 \langle L_{-\lambda}\mathcal{O}_3\left(\infty\right)\mathcal{O}_2(t)\mathcal{O}_1\left(0\right)\rangle=
 C\left(\Delta_3,\Delta_2,\Delta_1\right)
 \gamma_{\lambda}\left(\Delta_3,\Delta_2,\Delta_1\right)t^{|\lambda|},
 \end{eqnarray}
 where \cite{mmmjuly}
 \be\label{vertex}
 \gamma_{\lambda}\left(\Delta_1,\Delta_2,\Delta_3\right)=\prod_{j=1}^N\Bigl(\Delta_1-\Delta_3+\lambda_{j}\Delta_2+
 \sum_{k=1}^{j-1}\lambda_{k}\Bigr).
 \eb
 The action of $L_{-\lambda}$ on the field at infinity in (\ref{imprel}) should be understood as a result of successive
 contour integration with the energy-momentum tensor. This action can be transferred to the fields at $0$ and $t$
 by deformation of the contour.

 The relation (\ref{imprel}) is extremely important as it allows to determine the coefficients of the
 operator product expansion (OPE) of primary fields
 \be\label{mainope}
 \fl\mathcal{O}_2(t)\mathcal{O}_1(0)=\sum_{\alpha}\sum_{\mu\in\mathbb{Y}}
 C\left(\Delta_{\alpha},\Delta_2,\Delta_1\right)\beta_{\mu}\left(\Delta_{\alpha},\Delta_2,\Delta_{1}\right)
 t^{\Delta_{\alpha}-\Delta_1-\Delta_2+|\mu|}L_{-\mu}\mathcal{O}_{\alpha}\left(0\right).
 \eb
 Indeed, assuming orthonormality of the basis of primaries,
 $\langle\mathcal{O}_{\alpha}\left(\infty\right)\mathcal{O}_{\beta}\left(0\right)\rangle=\delta_{\alpha\beta}$, and
 considering the correlator of both sides of the last relation with the descendant $L_{-\lambda}\mathcal{O}_{\alpha}\left(\infty\right)$, one finds  that
 \be\label{coefsope}
 \beta_{\lambda}\left(\Delta_{\alpha},\Delta_2,\Delta_{1}\right)=\sum_{\mu\in\mathbb{Y}}
 \bigl[Q\left(\Delta_{\alpha}\right)\bigr]^{-1}_{\lambda\mu}\gamma_{\mu}\left(\Delta_{\alpha},\Delta_2,\Delta_{1}\right),
 \eb
 where $Q_{\lambda\mu}\left(\Delta_{\alpha}\right)=\langle L_{-\lambda}\mathcal{O}_{\alpha}\left(\infty\right)
 L_{-\mu}\mathcal{O}_{\alpha}\left(0\right)\rangle$ is the Kac-Shapovalov matrix. It can be computed algebraically
 as the matrix element of descendant states
 \be\label{shapo}
 Q_{\lambda\mu}\left(\Delta\right)=\langle\Delta|L_{\lambda_1}\ldots L_{\lambda_N}L_{-\mu_M}\ldots L_{-\mu_1}|\Delta\rangle,
 \eb
 where $|\Delta\rangle$ and $\langle\Delta|$ denote the highest weight vectors annihilated by all $L_{n>0}$ and, respectively, all
 $L_{n<0}$, satisfying $L_0|\Delta\rangle=\Delta|\Delta\rangle$, $\langle\Delta|L_0=\langle\Delta|\Delta$ and normalized
 as $\langle\Delta|\Delta\rangle=1$. It is easy to understand that $Q(\Delta)$ has a block-diagonal structure:
 $Q_{\lambda\mu}\left(\Delta\right)\sim \delta_{|\lambda|,|\mu|}$.

 We can now finally calculate the four-point correlator $\langle\mathcal{O}_4(\infty)\mathcal{O}_3(1)\mathcal{O}_2(t)\mathcal{O}_1(0)\rangle$.
 Replace therein the product of fields $\mathcal{O}_2(t)\mathcal{O}_1(0)$ by the OPE (\ref{mainope}) and
 then use (\ref{imprel0}) and (\ref{coefsope}). The result is
 \begin{eqnarray}\label{correl}
 \langle\mathcal{O}_4(\infty)\mathcal{O}_3(1)\mathcal{O}_2(t)\mathcal{O}_1(0)\rangle=\\
 \nonumber
 \fl \qquad =\sum_{\alpha}
 C\left(\Delta_4,\Delta_3,\Delta_{\alpha}\right)C\left(\Delta_{\alpha},\Delta_2,\Delta_{1}\right)
 t^{\Delta_{\alpha}-\Delta_1-\Delta_2}\mathcal{F}_c\left(\Delta_1,\Delta_2,\Delta_3,\Delta_4,\Delta_{\alpha};t\right),
 \end{eqnarray}
 where we have introduced the notation
 \begin{eqnarray}\label{CB}
 \fl\qquad \mathcal{F}_c\left(\Delta_1,\Delta_2,\Delta_3,\Delta_4,\Delta;t\right)=\sum_{\lambda,\mu\in\mathbb{Y}}
 \gamma_{\lambda}\left(\Delta,\Delta_3,\Delta_{4}\right)
 \bigl[Q\left(\Delta\right)\bigr]^{-1}_{\lambda\mu}
 \gamma_{\mu}\left(\Delta,\Delta_2,\Delta_{1}\right)t^{|\lambda|}.
 \end{eqnarray}
 The representation (\ref{correl}) separates model-dependent information (three-point
 functions $C\left(\Delta_i,\Delta_j,\Delta_k\right)$) from the universal pieces fixed solely by Virasoro
 symmetry.

  The function (\ref{CB}) is called \textit{four-point conformal block}. It is a power series in $t$ with
 coefficients depending on four external dimensions $\Delta_{1,2,3,4}$, one intermediate dimension~$\Delta$,
 and the central charge $c$ which enters via the Kac-Shapovalov matrix. These coefficients can in principle
 be calculated using (\ref{vertex}) and (\ref{shapo}). For the reader's convenience, we reproduce below
 several first terms of the series:
 \begin{eqnarray*}
 \mathcal{F}_c\left(\Delta_1,\Delta_2,\Delta_3,\Delta_4,\Delta;t\right)=
 1+\frac{\left(\Delta-\Delta_1+\Delta_2\right)\left(\Delta-\Delta_4+\Delta_3\right)}{2\Delta}\,t+\\
 \fl\qquad +\Biggl[\frac{\left(\Delta-\Delta_1+\Delta_2\right)\left(\Delta-\Delta_1+\Delta_2+1\right)
 \left(\Delta-\Delta_4+\Delta_3\right)\left(\Delta-\Delta_4+\Delta_3+1\right)}{2\Delta\left(1+2\Delta\right)}\,+\Biggr.\\
 \fl\qquad\Biggl.+\frac{\left(1+2\Delta\right)\left(\Delta_1+\Delta_2+\frac{\Delta\left(\Delta-1\right)-
 3\left(\Delta_1-\Delta_2\right)^2}{1+2\Delta}\right)\left(\Delta_4+\Delta_3+\frac{\Delta\left(\Delta-1\right)-
 3\left(\Delta_4-\Delta_3\right)^2}{1+2\Delta}\right)}{
 \left(1-4\Delta\right)^2+\left(c-1\right)\left(1+2\Delta\right)}\Biggr]\,\frac{t^2}{2}+\ldots
 \end{eqnarray*}

 Direct (i.e. based on (\ref{CB})) computation of conformal block coefficients  becomes rather complicated at higher levels. An explicit representation for arbitrary level was found only recently in a surprisingly different framework.

 \subsection{$\mathcal{N}=2$ SUSY theories}
 The AGT correspondence \cite{AGT} relates conformal blocks of 2D CFT to Nekrasov functions \cite{Nekrasov1,Nekrasov_Okounkov}.
 These functions represent the instanton parts of $\epsilon_1,\epsilon_2$-regularized partition functions in 4D $\mathcal{N}=2$
 SUSY quiver gauge theories.

 The simplest case of AGT correspondence deals with $SU(2)$ gauge theory with extra
 $N_f=2N_c=4$ fundamental (i.e. transforming in the spin-$\frac12$ representation of the gauge group) matter hypermultiplets
 with masses $\mu_1,\ldots,\mu_4$. Parameters of this theory are related to those of $4$-point conformal block
 on the sphere by
 \begin{eqnarray*}
 \mu_1=\alpha_3-\alpha_4+\frac{\epsilon}{2},\qquad &
 \mu_2=\alpha_1-\alpha_2+\frac{\epsilon}{2},\\
 \mu_3=\alpha_1+\alpha_2-\frac{\epsilon}{2},\qquad
 &\mu_4=\alpha_3+\alpha_{4}-\frac{\epsilon}{2},\\
 c=1+\frac{6\epsilon^2}{\epsilon_1\epsilon_2},\qquad & \epsilon=\epsilon_1+\epsilon_2, \\
 \Delta_{\nu}=\frac{\alpha_{\nu}\left(\epsilon-\alpha_{\nu}\right)}{\epsilon_1\epsilon_2},\qquad &\nu=1,2,3,4.
 \end{eqnarray*}
 The intermediate dimension $\Delta$ of conformal block is expressed via the
 eigenvalues $\pm a$ of the vacuum expectation value of scalar field in the gauge multiplet:
 \ben
 \Delta=\frac{\alpha\left(\epsilon-\alpha\right)}{\epsilon_1\epsilon_2},\qquad\quad\;\; \alpha=\frac{\epsilon}{2}+a.
 \ebn
 The parameter $t$ (anharmonic ratio of four points on the sphere) in the conformal block expansion is related to the
 bare complex coupling constant $\tau_{UV}$ on the gauge side by
 \ben
 t=\exp 2\pi i \tau_{UV},\qquad \tau_{UV}=\frac{4\pi i}{g_{UV}^2}+\frac{\theta_{UV}}{2\pi}.
 \ebn

 Partition function in the regularized theory is an integral over a compactified
 moduli space $\mathfrak{M}$ of instantons. The integral is given by a sum of explicitly computable contributions
 coming from fixed points of a torus action on $\mathfrak{M}$, which are labeled by pairs of partitions.
 On the CFT side, this is interpreted as an existence of a geometrically distinguished basis of states
 in the highest weight representations of the Virasoro algebra. For more details, generalizations and further references,
 the reader is referred to \cite{belavin}.

 When all four masses $\mu_{1,2,3,4}\rightarrow\infty$, the fundamental hypermultiplets decouple and we get pure gauge theory.
 Decoupling only some of them yields asymptotically free theories with $N_f<2N_c$.
 From the gauge theory point of view, the parameter $t_{N_{f}}$  can be considered as a dynamically generated scale.
 Taking into account the RG dependence of the coupling constant, one
 finds that in the decoupling process this scale should transform in the appropriate way:
 $\mu_{N_f}\to \infty$, $t_{N_f}\to 0$,
 $t_{N_f-1}=\mu_{N_f}t_{N_f}$ fixed \cite{Dorey,Seiberg}.
 The corresponding
 Nekrasov functions are related to irregular conformal blocks \cite{Gaiotto1,GT,MMM}.

 Conformal blocks relevant to Painlev\'e VI equation \cite{CFT_PVI} are characterized by the central charge $c=1$ and
 external dimensions $\theta_{\nu}^2$
 ($\nu=0,t,1,\infty$) so that we can set $\epsilon_1=-\epsilon_2=1$, $\alpha_1=\theta_0$, $\alpha_2=\theta_t$,
 $\alpha_3=\theta_1$, $\alpha_4=\theta_{\infty}$ and
 \begin{eqnarray*}
 \fl\qquad \qquad \mu_1=\theta_1-\theta_{\infty},\qquad \mu_2=\theta_0-\theta_t,\qquad \mu_3=\theta_0+\theta_t,\qquad
 \mu_4=\theta_1+\theta_{\infty}.
 \end{eqnarray*}
 Under such identification of parameters, the scaling limits corresponding to the 1st line of
 the coalescence scheme in Fig.~1 describe successive decoupling of the matter hypermultiplets:
  \begin{center}
 \begin{tikzpicture}[node distance=3cm, auto]
 \node (P6) {$\begin{array}{c} N_f=4 \\ (P_{_{\mathrm{VI}}})\end{array}$};
 \node (P5) [right of=P6] {$\begin{array}{c} N_f=3 \\ (P_{_{\mathrm{V}}})\end{array}$};
 \node (P31) [right of=P5] {$\begin{array}{c} N_f=2 \\ (P_{_{\mathrm{III_1}}})\end{array}$};
 \node (P32) [right of=P31] {$\begin{array}{c} N_f=1 \\ (P_{_{\mathrm{III_2}}})\end{array}$};
 \node (P33) [right of=P32, node distance=4cm] {$\begin{array}{c} \mathrm{pure\; gauge\; theory} \\ (P_{_{\mathrm{III_3}}})\end{array}$};
 \draw[->]  (P6) to node {\tiny $\mu_4\rightarrow\infty$} (P5);
  \draw[->] (P5) to node {\tiny $\mu_3\rightarrow\infty$} (P31);
   \draw[->] (P31) to node {\tiny $\mu_2\rightarrow\infty$} (P32);
    \draw[->] (P32) to node {\tiny $\mu_1\rightarrow\infty$} (P33);
 \end{tikzpicture}\\
 Fig. 3: Decoupling of matter hypermultiplets
 \end{center}
 This observation will be used in the next section for the construction of combinatorial series for
 $P_{_{\mathrm{V}}}$ and $P_{_{\mathrm{III_{1,2,3}}}}$ tau functions.

 \section{Solutions}\label{secsols}
 \subsection{Painlev\'e VI}
 Let us first recall the main result of \cite{CFT_PVI} as well as some motivation and evidence for it.
 \begin{conj}\label{conjpvi}
 Generic $P_{_{\mathrm{VI}}}$ tau function can be written in the form
 of conformal expansion around the critical point $t=0$:
 \be\label{tauvicexp}
 \fl
 \tau_{_{\mathrm{VI}}}(t)=\sum_{n\in\mathbb{Z}}C_{_{\mathrm{VI}}}\left(\theta_0,\theta_t,\theta_1,\theta_{\infty},\sigma+n\right)
 s_{_{\mathrm{VI}}}^n
 \,t^{\left(\sigma+n\right)^2-\theta_0^2-\theta_t^2}
 \mathcal{B}_{_{\mathrm{VI}}}\left(\theta_0,\theta_t,\theta_1,\theta_{\infty},\sigma+n;t\right).
 \eb
 The parameters $\sigma$ and $s_{_{\mathrm{VI}}}$ play the role of two integration constants, $\mathcal{B}_{_{\mathrm{VI}}}\left(\theta_0,\theta_t,\theta_1,\theta_{\infty},\sigma;t\right)$ coincides with
 conformal block function $\mathcal{F}_{c=1}\left(\theta_0^2,\theta_t^2,\theta_1^2,\theta_{\infty}^2,\sigma^2;t\right)$ and is explicitly
 given by combinatorial series
  \begin{eqnarray}\label{cbpvi}
 \fl\mathcal{B}_{_{\mathrm{VI}}}\left(\theta_0,\theta_t,\theta_1,\theta_{\infty},\sigma;t\right)=(1-t)^{2\theta_t\theta_1}\sum_{\lambda,\mu\in\mathbb{Y}}
 \mathcal{B}_{\lambda,\mu}^{^{\mathrm{(VI)}}}\left(\theta_0,\theta_t,\theta_1,\theta_{\infty},\sigma\right)   t^{|\lambda|+|\mu|},
 \end{eqnarray}
 \begin{eqnarray}\label{bpvi}
 \fl \mathcal{B}_{\lambda,\mu}^{^{\mathrm{(VI)}}}\left(\theta_0,\theta_t,\theta_1,\theta_{\infty},\sigma\right)&=
 \prod_{(i,j)\in\lambda}
 \frac{\left(\left(\theta_t+\sigma+i-j\right)^2-\theta_0^2\right)
 \left(\left(\theta_1+\sigma+i-j\right)^2-\theta_{\infty}^2\right)}{
 h_{\lambda}^2(i,j)\left(\lambda'_j+\mu_i-i-j+1+2\sigma\right)^2}\,\times\\
 \fl\nonumber &\times\prod_{(i,j)\in\mu}
 \frac{\left(\left(\theta_t-\sigma+i-j\right)^2-\theta_0^2\right)
 \left(\left(\theta_1-\sigma+i-j\right)^2-\theta_{\infty}^2\right)}{
 h_{\mu}^2(i,j)\left(\lambda_i+\mu'_j-i-j+1-2\sigma\right)^2}\,.
 \end{eqnarray}
 The structure constants in (\ref{tauvicexp}) are given by
 \begin{eqnarray}\label{scpvi}
 \fl C_{_{\mathrm{VI}}}\left(\theta_0,\theta_t,\theta_1,\theta_{\infty},\sigma\right)=\frac{\prod\limits_{\epsilon,\epsilon'=\pm}G\left[\begin{array}{c}
 1+\theta_t+\epsilon\theta_0+\epsilon'\sigma,1+\theta_1+\epsilon\theta_{\infty}+\epsilon'\sigma
 \end{array}\right]}{
 \prod_{\epsilon=\pm}G\left(1+2\epsilon\sigma\right)},
 \end{eqnarray}
 where $G\biggl[\begin{array}{c}\alpha_1,\ldots,\alpha_m \\ \beta_1,\ldots,\beta_n\end{array}\biggr]=\displaystyle \frac{\prod_{k=1}^m G\left(\alpha_k\right)}{\prod_{k=1}^n G\left(\beta_k\right)}$ and $G(z)$ denotes the Barnes function (see Appendix~A).
 \end{conj}

 The above claim was obtained in \cite{CFT_PVI} by identifying $\tau_{_{\mathrm{VI}}}(t)$ with a chiral correlator
 $\langle\mathcal{O}_{\mathcal{L}_{\infty}}(\infty)\mathcal{O}_{\mathcal{L}_1}(1)\mathcal{O}_{\mathcal{L}_t}(t)
 \mathcal{O}_{\mathcal{L}_0}(0)\rangle$ of Virasoro primary
 fields indexed by matrices $\mathcal{L}_{\nu}\in\mathfrak{sl}_2(\mathbb{C})$ which are related to monodromy matrices of
 the auxiliary linear problem for $P_{_{\mathrm{VI}}}$  by $\mathcal{M}_{\nu}=\exp 2\pi i \mathcal{L}_{\nu}\in SL\left(2,\mathbb{C}\right)$.
 The dimensions of $\mathcal{O}_{\mathcal{L}_{\nu}}$ are equal to $\Delta_{\nu}=\frac12\mathrm{Tr}\,\mathcal{L}_{\nu}^2=\theta_{\nu}^2$.

 Our main assumption is that this set of primaries closes under OPE algebra. The conservation of monodromy
 then implies that the dimension spectrum of primary fields appearing in the OPE of $\mathcal{O}_{\mathcal{L}_t}(t)\mathcal{O}_{\mathcal{L}_0}(0)$
 is discrete and has the form $\left(\sigma_{0t}+\mathbb{Z}\right)^2$,  where $2\cos2\pi\sigma_{0t}=\mathrm{Tr}\,\mathcal{M}_0\mathcal{M}_t$. This  fixes the structure of the $s$-channel expansion (\ref{tauvicexp}) upon identification $\sigma=\sigma_{0t}$. The constants $C_{_{\mathrm{VI}}}$ are obtained from  Jimbo's asymptotic formula \cite{jimbo} interpreted as a recursion relation,
 whereas (\ref{cbpvi})--(\ref{bpvi}) is nothing but the AGT representation for $c=1$ conformal block rewritten in
 terms of $P_{_{\mathrm{VI}}}$ parameters.

 In fact, Jimbo's formula also expresses the second integration constant $s_{_{\mathrm{VI}}}$ in terms of monodromy. To give
 an explicit
 relation, we need to introduce monodromy invariants
 \begin{eqnarray}
 p_{\nu}\;\,=2\cos2\pi\theta_{\nu}\;\;=\mathrm{Tr}\,\mathcal{M}_{\nu},\qquad\quad\;\;\nu=0,1,t,\infty,\\
 p_{\mu\nu}=2\cos2\pi\sigma_{\mu\nu}=\mathrm{Tr}\,\mathcal{M}_{\mu}\mathcal{M}_{\nu},\qquad \mu,\nu=0,t,1.
 \end{eqnarray}
 Similarly to the above, the quantities $\sigma_{1t}$ and $\sigma_{01}$ determine the spectrum
 of intermediate states in the $t$- and $u$-channel.
 The triple $\vec{\sigma}=\left(\sigma_{0t},\sigma_{1t},\sigma_{01}\right)$ provides the most symmetric way to label
 $P_{_{\mathrm{VI}}}$ transcendents with the same $\vec{\theta}=\left(\theta_0,\theta_t,\theta_1,\theta_{\infty}\right)$.
 The elements of this triple are not independent: they satisfy a constraint
 \begin{eqnarray}
 p_{0t}p_{1t}p_{01}+p_{0t}^2+p_{1t}^{2}+p_{01}^2-\omega_{0t}p_{0t}-\omega_{1t}p_{1t}-\omega_{01}p_{01}+\omega_4=4,
 \end{eqnarray}
 where
 \begin{eqnarray*}
 \omega_{0t}=p_0p_t+p_1p_{\infty},\\ \omega_{1t}=p_t p_1+p_0 p_{\infty},\\
 \omega_{01}=p_0p_1+ p_t p_{\infty},\\
 \omega_4=p_0^2+p_t^2+p_1^2+p_{\infty}^2+p_0 p_t p_1 p_{\infty}.
 \end{eqnarray*}
 Hence, for fixed $\sigma_{0t}$, $\sigma_{1t}$ there are at most two possible values for $p_{01}$.

 Now $s_{_{\mathrm{VI}}}$ can be written as
 \be\label{spvi}
 s_{_{\mathrm{VI}}}=\frac{\left(p_{1t}'-p_{1t}\right)-\left(p_{01}'-p_{01}\right)e^{2\pi i \sigma_{0t}}}{\left(2\cos2\pi\left(\theta_t-\sigma_{0t}\right)-p_0\right)
 \left(2\cos2\pi\left(\theta_1-\sigma_{0t}\right)-p_{\infty}\right)},
 \eb
 where we have introduced the notation
 \begin{eqnarray*}
  p_{0t}'\,=\omega_{0t}\,-p_{0t}-p_{1t}p_{01},\\
  p_{1t}'\,=\omega_{1t}\,-p_{1t}-p_{0t}p_{01},\\
 p_{01}'=\omega_{01}-p_{01}-p_{0t}p_{1t}.
 \end{eqnarray*}
 \begin{rmk}\label{expar1} Combinatorial expansions of type (\ref{tauvicexp}) can also be found around the
 two remaining critical points $t=1,\infty$, as their role is completely analogous to that of $t=0$.
 For instance, the  series around $t=1$ is obtained  by the exchange \cite{jimbo,dyson2f1}
 \be
 t\leftrightarrow 1-t,\qquad \theta_0\leftrightarrow \theta_1,\qquad \sigma_{0t}\leftrightarrow \sigma_{1t},\qquad
 p_{01}\leftrightarrow p_{01}'.
 \eb
 This gives
 \begin{eqnarray}\label{tauvicexp2}
 \chi_{01}\bigl(\vec{\theta},\vec{\sigma}\bigr)\,\tau_{_{\mathrm{VI}}}(t)=\\
 \fl\nonumber =\sum_{n\in\mathbb{Z}}C_{_{\mathrm{VI}}}\left(\theta_1,\theta_t,\theta_0,\theta_{\infty},\sigma_{1t}+n\right)
 \tilde{s}_{_{\mathrm{VI}}}^n
 \left(1-t\right)^{\left(\sigma_{1t}+n\right)^2-\theta_t^2-\theta_1^2}
 \mathcal{B}_{_{\mathrm{VI}}}\left(\theta_1,\theta_t,\theta_0,\theta_{\infty},\sigma_{1t}+n;1-t\right),
 \end{eqnarray}
 with $\mathcal{B}_{_{\mathrm{VI}}}$ and $C_{_{\mathrm{VI}}}$ defined in (\ref{cbpvi})--(\ref{scpvi}) and
 \be\label{stpvi}
  \tilde{s}_{_{\mathrm{VI}}}=\frac{\left(p_{0t}'-p_{0t}\right)-\left(p_{01}'-p_{01}\right)e^{-2\pi i \sigma_{1t}}}{\left(2\cos2\pi\left(\theta_t-\sigma_{1t}\right)-p_1\right)
 \left(2\cos2\pi\left(\theta_0-\sigma_{1t}\right)-p_{\infty}\right)}.
 \eb
 Since the normalization of $\tau_{_{\mathrm{VI}}}(t)$ is already implicitly fixed by (\ref{tauvicexp}),
 the expansion (\ref{tauvicexp2}) contains an additional overall constant factor $\left[\chi_{01}\bigl(\vec{\theta},\vec{\sigma}\bigr)\right]^{-1}$.
 Finding explicit form of this connection coefficient  is an important open problem which will be treated in a separate paper. Note, however, that $\chi_{01}\bigl(\vec{\theta},\vec{\sigma}\bigr)$ disappears
 from $P_{_{\mathrm{VI}}}$ functions
 $\sigma_{_{\mathrm{VI}}}(t)$ and $q_{_{\mathrm{VI}}}(t)$.
 \end{rmk}

 Painlev\'e VI equation (\ref{sigpvi}) allows to compute the tau function expansions near the critical points recursively, order by order,
 starting from the leading asymptotic terms determined by Jimbo's formula. Comparing the result with Conjecture~\ref{conjpvi} provides
 the most straight\-forward and convincing test of the latter. Keeping
 all $\vec{\theta}$, $\vec{\sigma}$ arbitrary, we have checked in this way (see Section~3 of \cite{CFT_PVI}
 for the details) about 30 first terms of the
 asymptotic expansion.
  Also, in a few special cases where $P_{_{\mathrm{VI}}}$ solutions are known
 explicitly, the check can be carried out to arbitrary order. This includes
 Picard elliptic solutions and the simplest solutions of Riccati/Chazy type, which correspond to
 Ashkin-Teller conformal blocks \cite{Zamo_AT} and correlators involving low-level degenerate fields.
 More complicated Riccati solutions are discussed in Subsection~\ref{Gessel} of the present paper.

 \begin{figure}[!h]
 \begin{center}
 \resizebox{15cm}{!}{
 \includegraphics{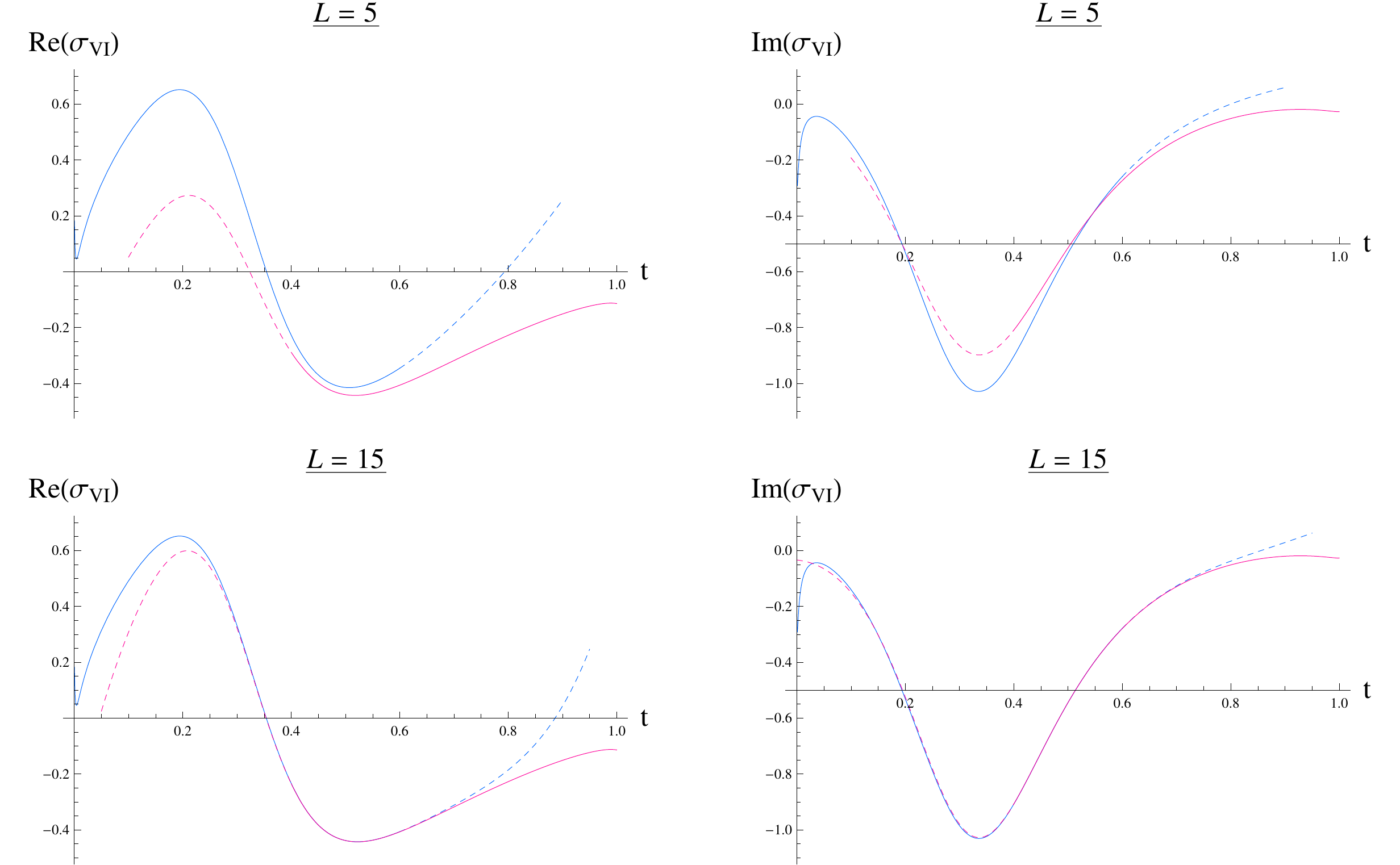}} \\
 Fig. 4: Truncated $P_{_{\mathrm{VI}}}$ series at $t=0$ and $t=1$ with $\vec{\theta}=\left( 0.1902 + 0.3106 i , 0.4182 - 0.2109 i ,
 0.3429 + 0.3314 i , 0.0163 + 0.1805 i \right)$, $\vec{\sigma}=\left(-0.3272 - 0.4811 i, 0.0958 + 0.3168 i,
 0.4762 + 0.1028 i\right)$
 \end{center}
 \end{figure}

 Numerical efficiency of the expansions (\ref{tauvicexp}), (\ref{tauvicexp2}) is illustrated in Fig.~4.
 For random complex $\vec{\theta}$, $\vec{\sigma}$ we plot on the same graph the series
 for $\sigma_{_{\mathrm{VI}}}(t)$ around $t=0$ (blue line) and $t=1$ (red line) keeping the terms up to $O\left(t^{L}\right)$
 and $O\left(\left(1-t\right)^{L}\right)$ with $L=5,15$. Zooming near the endpoints $t=0,1$ would display
 oscillations of rapidly increasing frequency and decreasing amplitude due to non-zero imaginary parts
 of $\sigma_{0t}$, $\sigma_{1t}$.

 \subsection{Painlev\'e V and III's}
 Next we consider the scaling limit $P_{_{\mathrm{VI}}}\rightarrow P_{_{\mathrm{V}}}$ given by
 (\ref{limpvitopv1})--(\ref{limpvitopv4}).
 Conformal block function $\mathcal{B}_{_{\mathrm{VI}}}\left(\theta_0,\theta_t,\frac{\Lambda+\theta_*}{2},\frac{\Lambda-\theta_*}{2},\sigma;\frac{t}{\Lambda}\right)$
 has a well-defined limit as $\Lambda\rightarrow\infty$, which can be calculated termwise in (\ref{cbpvi})--(\ref{bpvi}).
 The asymptotics of the structure constants $C_{_{\mathrm{VI}}}\left(\theta_0,\theta_t,\frac{\Lambda+\theta_*}{2},\frac{\Lambda-\theta_*}{2},\sigma\right)$ ensures
 consistency of the expansion (\ref{tauvicexp}) with the limit (\ref{limpvitopv4}). More precisely,
 using the estimate (\ref{barnesaux0}) from the
 Appendix~A, it is easy to check that
 \be\label{strpvlim}
 \lim_{\Lambda\rightarrow\infty}
 \frac{{\Lambda}^{-\sigma^2}
 C_{_{\mathrm{VI}}}\left(\theta_0,\theta_t,\frac{\Lambda+\theta_*}{2},\frac{\Lambda-\theta_*}{2},\sigma\right)}{
 G^2\left(1+\Lambda\right)}=C_{_{\mathrm{V}}}\left(\theta_0,\theta_t,\theta_*,\sigma\right),
 \eb
 where
  \be\label{strpvv2}
 \fl \qquad C_{_{\mathrm{V}}}\left(\theta_0,\theta_t,\theta_*,\sigma\right)=
 \prod_{\epsilon=\pm}G\biggl[\begin{array}{c}
 1+\theta_*+\epsilon\sigma,1+\theta_t+\theta_0+\epsilon\sigma,1+\theta_t-\theta_0+\epsilon\sigma \\
 1+2\epsilon\sigma
 \end{array}\biggr].
 \eb
 One could even completely get rid of the denominator in the l.h.s. of (\ref{strpvlim}) by modifying
 the normalization of $P_{_{\mathrm{VI}}}$ tau function in (\ref{tauvicexp}) (e.g. by dividing all structure constants
 in (\ref{scpvi}) by a $\sigma$-independent factor
 $G^2\left(1+\theta_1+\theta_{\infty}\right)$).

 Altogether, this leads to
 \begin{conj}\label{conjpv}
 $P_{_{\mathrm{V}}}$ is solved by the following tau function expansion at $t=0$:
 \be\label{tauvcexp}
 \tau_{_{\mathrm{V}}}(t)=\sum_{n\in\mathbb{Z}}C_{_{\mathrm{V}}}\left(\theta_0,\theta_t,\theta_*,\sigma+n\right)s_{_{\mathrm{V}}}^n
 \,t^{\left(\sigma+n\right)^2}
 \mathcal{B}_{_{\mathrm{V}}}\left(\theta_0,\theta_t,\theta_*,\sigma+n;t\right).
 \eb
 Here again $\sigma$ and $s_{_{\mathrm{V}}}$ are arbitrary parameters, irregular conformal block $\mathcal{B}_{_{\mathrm{V}}}\left(\theta_0,\theta_t,\theta_*,\sigma;t\right)$
 is a power series defined by
 \begin{eqnarray}\label{CBPV1}
 \mathcal{B}_{_{\mathrm{V}}}\left(\theta_0,\theta_t,\theta_*,\sigma;t\right)=e^{-\theta_t t}\sum_{\lambda,\mu\in\mathbb{Y}}
 \mathcal{B}_{\lambda,\mu}^{^{\mathrm{(V)}}}\left(\theta_0,\theta_t,\theta_*,\sigma\right)   t^{|\lambda|+|\mu|},
 \end{eqnarray}
  \begin{eqnarray}\label{CBPV2}
 \mathcal{B}_{\lambda,\mu}^{^{\mathrm{(V)}}}\left(\theta_0,\theta_t,\theta_*,\sigma\right)&=
 \prod_{(i,j)\in\lambda}
 \frac{\left(\theta_*+\sigma+i-j\right)\left(\left(\theta_t+\sigma+i-j\right)^2-\theta_0^2\right)}{
 h_{\lambda}^2(i,j)\left(\lambda'_j+\mu_i-i-j+1+2\sigma\right)^2}\,\times\\
 \nonumber&\times \prod_{(i,j)\in\mu}
 \frac{\left(\theta_*-\sigma+i-j\right)\left(\left(\theta_t-\sigma+i-j\right)^2-\theta_0^2\right)}{
 h_{\mu}^2(i,j)\left(\lambda_i+\mu'_j-i-j+1-2\sigma\right)^2}\,,
 \end{eqnarray}
 and the structure constants $C_{_{\mathrm{V}}}\left(\theta_0,\theta_t,\theta_*,\sigma\right)$ are given by (\ref{strpvv2}).
 \end{conj}

 The second $P_{_{\mathrm{V}}}$ critical point  $t=\infty$ corresponds to irregular singularity
 of the associated $2\times 2$ linear system and is obtained by the fusion of two $P_{_{\mathrm{VI}}}$ critical points
 $1,\infty$. The  expansion around this point cannot be extracted from $P_{_{\mathrm{VI}}}$ series
 and requires the knowledge of complete irregular OPEs. For the same reason, we are so far unable to treat
 $P_{_{\mathrm{IV}}}$, $P_{_{\mathrm{II}}}$ and $P_{_{\mathrm{I}}}$. However, long-distance expansions of this kind are available
 in a few special cases where the solutions of $P_{_{\mathrm{V,III}}}$ can be expressed in terms of Fredholm determinants, see Section~\ref{secapp}.

  Because of the presence of irregular singular points, monodromy data for $P_{_{\mathrm{V}}}$ involve Stokes multipliers.
  The expression for the integration constants $\sigma$, $s_{_{\mathrm{V}}}$ of Conjecture~\ref{conjpv}  in terms of monodromy can be extracted
  from Jimbo's paper \cite{jimbo}.

 Repeating the previous arguments almost literally for the scaling limits (\ref{pvtopiii1first})--(\ref{pvtopiii1last}),
 (\ref{piii1topiii2first})--(\ref{piii1topiii2last}) and  (\ref{piii2topiii3first})--(\ref{piii2topiii3last}), one obtains
 short-distance expansions for tau functions of three nontrivial $P_{_{\mathrm{III}}}$ equations:
  \begin{conj}\label{conjp31}
 Expansion of $\tau_{_{\mathrm{III}'_1}}(t)$ at $t=0$ can be written as
 \be\label{tauexpp3v1}
 \tau_{_{\mathrm{III}'_1}}(t)=\sum_{n\in\mathbb{Z}}C_{_{\mathrm{III}'_1}}\left(\theta_*,\theta_{\star},\sigma+n\right)
 s_{_{\mathrm{III}'_1}}^n\,
 t^{\left(\sigma+n\right)^2}
 \mathcal{B}_{_{\mathrm{III}'_1}}\left(\theta_*,\theta_{\star},\sigma+n;t\right),
 \eb
 where the irregular conformal block $\mathcal{B}_{_{\mathrm{III}'_1}}\left(\theta_*,\theta_{\star},\sigma;t\right)$
 is given by
 \begin{eqnarray}
 \label{cb31exp}
 \mathcal{B}_{_{\mathrm{III}'_1}}\left(\theta_*,\theta_{\star},\sigma;t\right)=e^{-\frac{t}{2}}\sum_{\lambda,\mu\in\mathbb{Y}}
 \mathcal{B}_{\lambda,\mu}^{^{\mathrm{(III'_1)}}}\left(\theta_*,\theta_{\star},\sigma\right)   t^{|\lambda|+|\mu|},
 \end{eqnarray}
  \begin{eqnarray}\label{cb31coef}
 \mathcal{B}_{\lambda,\mu}^{^{\mathrm{(III'_1)}}}\left(\theta_*,\theta_{\star},\sigma\right)&=
 \prod_{(i,j)\in\lambda}
 \frac{\left(\theta_*+\sigma+i-j\right)\left(\theta_{\star}+\sigma+i-j\right)}{
 h_{\lambda}^2(i,j)\left(\lambda'_j+\mu_i-i-j+1+2\sigma\right)^2} \times \\
 \nonumber &\times\prod_{(i,j)\in\mu}
 \frac{\left(\theta_*-\sigma+i-j\right)\left(\theta_{\star}-\sigma+i-j\right)}{
 h_{\mu}^2(i,j)\left(\lambda_i+\mu'_j-i-j+1-2\sigma\right)^2}\,,
 \end{eqnarray}
 and the structure constants can be written as
 \be\label{cp3v1}
 C_{_{\mathrm{III}'_1}}\left(\theta_*,\theta_{\star},\sigma\right)=
 \prod_{\epsilon=\pm}G\biggl[\begin{array}{c}1+\theta_*+\epsilon\sigma,1+\theta_{\star}+\epsilon\sigma \\ 1+2\epsilon\sigma \end{array}\biggr].
 \eb
 \end{conj}

  \begin{conj}
 Expansion of $\tau_{_{\mathrm{III}'_2}}(t)$ at $t=0$ is given by
 \be\label{taup32cexp}
 \tau_{_{\mathrm{III}'_2}}(t)=\sum_{n\in\mathbb{Z}}C_{_{\mathrm{III}'_2}}\left(\theta_*,\sigma+n\right)s_{_{\mathrm{III}'_2}}^n
 t^{\left(\sigma+n\right)^2}
 \mathcal{B}_{_{\mathrm{III}'_2}}\left(\theta_*,\sigma+n;t\right),
 \eb
 with arbitrary $\sigma$, $s_{_{\mathrm{III}'_2}}$ and
 \begin{eqnarray}
 \mathcal{B}_{_{\mathrm{III}'_2}}\left(\theta_*,\sigma;t\right)=\sum_{\lambda,\mu\in\mathbb{Y}}
 \mathcal{B}_{\lambda,\mu}^{^{\mathrm{(III'_2)}}}\left(\theta_*,\sigma\right)   t^{|\lambda|+|\mu|},
 \end{eqnarray}
  \begin{eqnarray}
 \mathcal{B}_{\lambda,\mu}^{^{\mathrm{(III'_2)}}}\left(\theta_*,\sigma\right)&=
 \prod_{(i,j)\in\lambda}
 \frac{\theta_*+\sigma+i-j}{
 h_{\lambda}^2(i,j)\left(\lambda'_j+\mu_i-i-j+1+2\sigma\right)^2}\times \\
 \nonumber & \times \prod_{(i,j)\in\mu}
 \frac{\theta_*-\sigma+i-j}{
 h_{\mu}^2(i,j)\left(\lambda_i+\mu'_j-i-j+1-2\sigma\right)^2},
 \end{eqnarray}
 \begin{eqnarray}
 C_{_{\mathrm{III}'_2}}\left(\theta_*,\sigma\right)=
 \prod_{\epsilon=\pm}
 \frac{G\bigl(1+\theta_*+\epsilon\sigma\bigr)}{
 G\bigl(1+2\epsilon\sigma\bigr)}.
 \end{eqnarray}
 \end{conj}
   \begin{conj}\label{conjpiii3}
 Expansion of $P_{_{\mathrm{III}'_3}}$ tau function at $t=0$ is:
 \be \label{piii3first}
 \tau_{_{\mathrm{III}'_3}}(t)=\sum_{n\in\mathbb{Z}}C_{_{\mathrm{III}'_3}}\left(\sigma+n\right)s_{_{\mathrm{III}'_3}}^n
 t^{\left(\sigma+n\right)^2}
 \mathcal{B}_{_{\mathrm{III}'_3}}\left(\sigma+n;t\right),
 \eb
 where
 \begin{eqnarray}
 \mathcal{B}_{_{\mathrm{III}'_3}}\left(\sigma;t\right)=\sum_{\lambda,\mu\in\mathbb{Y}}
 \mathcal{B}_{\lambda,\mu}^{^{\mathrm{(III'_3)}}}\left(\sigma\right)   t^{|\lambda|+|\mu|},
 \end{eqnarray}
  \begin{eqnarray}
 \mathcal{B}_{\lambda,\mu}^{^{\mathrm{(III'_3)}}}\left(\sigma\right)&=
 \biggl[\prod_{(i,j)\in\lambda}
 h_{\lambda}(i,j)\left(\lambda'_j+\mu_i-i-j+1+2\sigma\right) \biggr.\times  \\
 \nonumber &\;\;\times\biggl. \prod_{(i,j)\in\mu}
 h_{\mu}(i,j)\left(\lambda_i+\mu'_j-i-j+1-2\sigma\right)\biggr]^{-2}.
 \end{eqnarray}
 \begin{eqnarray}
 \label{piii3last}
 C_{_{\mathrm{III}'_3}}\left(\sigma\right)=\bigl[G\left(1+2\sigma\right)G\left(1-2\sigma\right)\bigr]^{-1}.
 \end{eqnarray}
 \end{conj}
 \noindent As in the $P_{_{\mathrm{VI}}}$ case,  Conjectures~\ref{conjpv}--\ref{conjpiii3} can be verified by
 iterative reconstruction of the tau function expansions from the leading asymptotic terms
 using the  equations $P_{_{\mathrm{V,III'_{1,2,3}}}}$.

 \subsection{Classical solutions: AGT vs Gessel's theorem}\label{Gessel}
 In 2002, Forrester and Witte \cite{FW} have proved a remarkable determinant representation for
 a family of Riccati solutions of $P_{_{\mathrm{VI}}}$. Their result can be restated as follows.
 Define a five-parameter family of $N\times N$ Toeplitz determinants
 \begin{eqnarray}\label{toepdet}
  D_N^{(\nu,\nu',\eta,\xi)}(t)=\mathrm{det}\left[A^{(\nu,\nu',\eta,\xi)}_{j-k}(t)\right]_{j,k=0}^{N-1},\\
 \nonumber
 \fl\qquad A^{(\nu,\nu',\eta,\xi)}_{m}(t)=
 \frac{\Gamma\left(1+\nu'\right)t^{\frac{\eta-m}{2}}\left(1-t\right)^{\nu}}{\Gamma\left(1+\eta-m\right)\Gamma\left(1-\eta+m+\nu'\right)} \,{}_{2}F_1\biggl[\begin{array}{c} -\nu,1+\nu'\\ 1+\eta-m\end{array}\biggr|\biggl. \frac{t}{t-1}\biggr]+\\
 \label{af21}
 \fl\qquad\qquad\qquad\quad +\frac{\xi\Gamma\left(1+\nu\right)t^{\frac{m-\eta}{2}}\left(1-t\right)^{\nu'}}{\Gamma\left(1-\eta+m\right)\Gamma\left(1+\eta-m+\nu\right)} \;\,{}_{2}F_1\biggl[\begin{array}{c} 1+\nu,-\nu'\\ 1-\eta+m\end{array}\biggr|\biggl. \frac{t}{t-1}\biggr].
 \end{eqnarray}
 Then the function
 \be\label{taufw}
 \tau_N^{(\nu,\nu',\eta,\xi)}(t)=\left(1-t\right)^{-\frac{N\left(N+\nu+\nu'\right)}{2}} D_N^{(\nu,\nu',\eta,\xi)}(t)
 \eb
 is a tau function of $P_{_{\mathrm{VI}}}$ with parameters
 \ben
 \left(\theta_0,\theta_t,\theta_1,\theta_{\infty}\right)_{_{\mathrm{VI}}}=\frac{1}{2}\left(\eta,N,-N-\nu-\nu',\nu-\nu'+\eta\right).
 \ebn

 Looking at the asymptotic expansions of $D_N^{(\nu,\nu',\eta,\xi)}(t)$ at $0$ and $1$, one can also identify
 the monodromy exponents
 \ben
 \left(\sigma_{0t},\sigma_{1t},\sigma_{01}\right)_{_{\mathrm{VI}}}=\frac12\left(N+\eta,\nu+\nu',N+\nu-\nu'+\eta\right).
 \ebn
 Almost all structure constants in (\ref{tauvicexp}) vanish because of the relations $\theta_t=\frac{N}{2}$, $\sigma_{0t}=\theta_0+\theta_t$ (recall that Barnes $G$-function has zeros at negative integer values of the argument).
 The only non-zero constants correspond to $n=0,-1,\ldots,-N$, so that
 there remain only $N+1$ conformal blocks.
 The parameter $s_{_{\mathrm{VI}}}$ in (\ref{tauvicexp}) is related to $\xi$ in (\ref{af21}) by
 \be\label{xis}
 \xi s_{_{\mathrm{VI}}}=\frac{\sin\pi\nu\sin\pi(\eta-\nu')}{\sin\pi\nu'\sin\pi(\eta+\nu)}.
 \eb

 Let us now consider in more detail the case $\xi\rightarrow 0$. Then (\ref{xis}) implies
 that $s_{_{\mathrm{VI}}}\rightarrow\infty$, which means that
 the expansion (\ref{tauvicexp}) at $t=0$ contains only one ($n=0$) conformal block
 $\mathcal{B}_{_{\mathrm{VI}}}\left(\frac{\eta}{2},\frac{N}{2},\frac{N+\nu+\nu'}{2},\frac{\nu-\nu'+\eta}{2},\frac{N+\eta}{2};t\right)$.
 The product over boxes of $\mu$ in the AGT representation (\ref{bpvi}) contains a factor $i-j$ due to the relation $\sigma_{0t}=\theta_0+\theta_t$. Since this expression vanishes for the
 box $(1,1)$, the quantity $\mathcal{B}_{\lambda,\mu}^{^{\mathrm{(VI)}}}$ does so for any non-empty $\mu$.
 Moreover, the factor $i-j+N$ in the product over boxes of $\lambda$ reduces the summation in (\ref{cbpvi}) to Young diagrams
 with $\lambda_1\leq N$ (i.e. with the length of their first row not exceeding $N$). Therefore, Conjecture~\ref{conjpvi}
 for the above parameters is equivalent to the following identity:
 \begin{eqnarray}\label{gesseldef}
 \fl D_N^{(\nu,\nu',\eta,0)}(t)=C_N\!\!\!\!\!\sum_{\lambda\in\mathbb{Y}|\lambda_1\leq N}t^{|\lambda|+\frac{N\eta}{2}}\prod_{(i,j)\in\lambda}
 \frac{i-j+N}{i-j+N+\eta}\frac{\left(i-j-\nu\right)\left(i-j-\nu'+\eta\right)}{h_{\lambda}^2(i,j)},
 \end{eqnarray}
 where the constant prefactor
 \be\label{prefac}
 C_N=G\biggl[\begin{array}{c}1+N,1+\nu'+N,1+\eta,1-\eta+\nu'\\
 1+\eta+N,1-\eta+\nu'+N,1+\nu'\end{array}\biggr]
 \eb
 can be computed using pure Fisher-Hartwig determinant.

 In the limit $\eta\rightarrow 0$, the left hand side of (\ref{gesseldef}) reduces to $N\times N$ Toeplitz determinant with the symbol
 \be\label{toep}
 A(\zeta)=\left(1+\sqrt{t}\,\zeta\right)^{\nu}\left(1+\sqrt{t}\,\zeta^{-1}\right)^{\nu'}.
 \eb
 Also, $C_N=1$ and the first factor in the product on the right  disappears so that the r.h.s. coincides with
 the length distribution
 function of the first row of a random Young diagram distributed according to the so-called $z$-measure \cite{borodin}.
 The equality (\ref{gesseldef}) can then be rigorously demonstrated using a dual version of Gessel's theorem
 \cite{gessel,twgessel}.
 \begin{rmk}
 We draw the reader's attention to the fact that Toeplitz determinant with the symbol (\ref{toep})
 with $\nu=-\nu'=\frac12$ coincides with diagonal two-point Ising spin correlation function  on the infinite
 square lattice. Its relation to $P_{_{\mathrm{VI}}}$ is rather well-known \cite{jm81}. It is intriguing, however,
 that this lattice correlator is equal to a (particular limit of) conformal block in continuous
 2D CFT with $c=1$.
 \end{rmk}

 Analogous results  for $P_{_{\mathrm{V}}}$ and $P_{_{\mathrm{III'_1}}}$ can be obtained by successively sending
 $\nu'$ and $\nu$ to infinity. For instance, consider instead of
 $A^{(\nu,\nu',\eta,\xi)}(t)$ and $\tau_N^{(\nu,\nu',\eta,\xi)}(t)$
 the quantities
 \begin{eqnarray}
 \label{toep1f1}
 A^{(\nu,\eta,\xi)}_m(t)=\frac{t^{\frac{\eta-m}{2}}}{\Gamma\left(1+\eta-m\right)}\,{}_1F_1\left(-\nu,1+\eta-m,-t\right)+\\
 \nonumber\qquad\qquad +\frac{\xi\Gamma\left(1+\nu\right)t^{\frac{m-\eta}{2}}e^{-t}}{
 \Gamma\left(1-\eta+m\right)\Gamma\left(1+\eta-m+\nu\right)}
 \,{}_1F_1\left(1+\nu,1-\eta+m,t\right),\\
 \label{toeptau1f1}\tau_N^{(\nu,\eta,\xi)}(t)=t^{\frac{N^2+\eta^2}{4}}e^{\frac{Nt}{2}}\;
 \mathrm{det}\left[A^{(\nu,\eta,\xi)}_{j-k}(t)\right]_{j,k=0}^{N-1}
 \end{eqnarray}
 then $\tau_N^{(\nu,\eta,\xi)}(t)$ is a tau function of $P_{_{\mathrm{V}}}$
 with $\left(\theta_0,\theta_t,\theta_*\right)_{_{\mathrm{V}}}=\frac{1}{2}\left(\eta,N,N+\eta+2\nu\right)$. Similarly, if we
 define
   \begin{eqnarray}
 \label{toepbessel}A^{(\eta,\xi)}_m(t)=I_{\eta-m}\left(2\sqrt{t}\right)+\xi I_{m-\eta}\left(2\sqrt{t}\right),\\
 \label{toeptaubessel}\tau_N^{(\eta,\xi)}(t)=t^{\frac{N^2+\eta^2}{4}}e^{-\frac{t}{2}}\;\mathrm{det}\left[A^{(\eta,\xi)}_{j-k}(t)\right]_{j,k=0}^{N-1},
 \end{eqnarray}
 then $\tau_N^{(\eta,\xi)}(t)$ is a $P_{_{\mathrm{III'_1}}}$ tau function  with $\theta_*=\frac{N+\eta}{2}$,
 $\theta_{\star}=\frac{N-\eta}{2}$. For $\xi=0$ and $\eta\rightarrow0$ the symbols of Toeplitz determinants
 (\ref{toeptau1f1}), (\ref{toeptaubessel}) are smooth and can be written as $\left(1+\sqrt{t}\,\zeta\right)^{\nu}e^{\sqrt{t}\,\zeta^{-1}}$ and $e^{ \sqrt{t}\left(\zeta+\zeta^{-1}\right)}$. Gessel representations of these determinants coincide with
 the results derived from Conjectures~\ref{conjpv} and~\ref{conjp31}.

 In the general case $\xi\neq 0$, the function $\tau^{(\nu,\nu',\eta,\xi)}(t)$ is a polynomial of degree
 $N$ in $\xi$. The coefficients of $N+1$ different powers of $\xi$ are $s$-channel conformal blocks  with internal dimensions
 $\left(\theta_0+\theta_t-k\right)^2$, where $k=0,\ldots,N$. Alternatively,
 one can first transform
 hypergeometric functions to make them depend on $1-t$ and then expand the determinant in powers of $\tilde \xi$,
 the analog of parameter $\xi$.
 The result has the form (\ref{tauvicexp2}) of a sum of $t$-channel conformal blocks with internal dimensions $\left(\theta_1+\theta_t-k\right)^2$,
 again with $k=0,\ldots,N$. The relations between the expansion parameters are given by (\ref{xis}),
$\tilde \xi  \tilde{s}_{_{\mathrm{VI}}}= \xi  {s}_{_{\mathrm{VI}}}={\cal K}$ and $(1-{s}_{_{\mathrm{VI}}})(1-\tilde{s}_{_{\mathrm{VI}}})=1+{\cal K}$.

 The CFT interpretation of this picture is as follows. The tau function (\ref{taufw}) is a four-point correlator of primaries
 which involves level~$N+1$ degenerate field (here $\mathcal{O}_{\mathcal{L}_t}(t)$). Its expansions at $t=0$ and $t=1$ incorporate all
 allowed intermediate dimensions. Determinant representation (\ref{toepdet}) can in fact be used to compute the fusion matrix
 for the corresponding two sets of conformal blocks. This task simplifies in the case $\xi=0$, where we are left with one $s$-channel block
 tranforming into a linear combination of $N+1$ $t$-channel ones.

 \section{Examples and applications}
 \label{secapp}
 \subsection{Integrable kernels}
 In many applications of Painlev\'e equations the relevant tau functions
 can be written as Fredholm determinants of scalar integral operators of the form $\mathrm{det}\left(1-K|_I\right)$, where
 $K|_I$ denotes the restriction of the kernel $K(x,y)$ to some interval $I\subset\mathbb{R}$. These kernels
 usually have integrable form, that~is
 \be\label{intform}
 K(x,y)=\lambda\,\frac{\varphi(x)\psi(y)-\psi(x)\varphi(y)}{x-y},\qquad \lambda\in\mathbb{C}.
 \eb
 As is well-known,  given $I=\bigcup_{j=1}^{2n}\left(a_{2j-1},a_{2j}\right)$ and
  $\varphi$, $\psi$ verifying  the differentiation formulas
 \ben
 \left(\begin{array}{l}\varphi'(x) \\ \psi'(x) \end{array}\right)=A(x)\left(\begin{array}{l}\varphi(x) \\ \psi(x) \end{array}\right),
 \ebn
 with some rational matrix $A(x)$, the corresponding Fredholm determinant satisfies a system of PDEs with respect to
 $\left\{a_j\right\}$  \cite{tweqs}. For
 $\varphi$, $\psi$ given by classical special functions and  sufficiently simple $I$, this system
 can often be solved in terms of Painlev\'e functions \cite{Forrester_book,dyson2f1,twairy,twbessel,tweqs,FWBessel}.

 \subsubsection{Hypergeometric kernel.} The most general known example corresponds to the choice
 \begin{eqnarray*}
 \fl\varphi_{_{\mathrm{G}}}(x)=&\,\Gamma\biggl[\begin{array}{c}
 1+\nu+\eta,1+\nu'+\eta'\\
 2+\nu+\nu'+\eta+\eta'\end{array}\biggr]\frac{x^{\frac{2+\nu+\nu'+\eta+\eta'}{2}}}{
 (1-x)^{\frac{2+\nu+\nu'+2\eta'}{2}}}\,{}
 _2F_1\biggl[\begin{array}{c}1+\nu+\eta',1+\nu'+\eta' \\ 2+\nu+\nu'+\eta+\eta'\end{array}\biggr|\frac{x}{x-1}\biggr],\\
 \fl\psi_{_{\mathrm{G}}}(x)=&\,
 \Gamma\biggl[\begin{array}{c}
 1+\nu+\eta',1+\nu'+\eta\\
 1+\nu+\nu'+\eta+\eta'\end{array}\biggr]
 \frac{x^{\frac{\nu+\nu'+\eta+\eta'}{2}}}{
 (1-x)^{\frac{\nu+\nu'+2\eta'}{2}}}\,{}
 _2F_1\biggl[\begin{array}{c}\nu+\eta',\nu'+\eta' \\ \nu+\nu'+\eta+\eta'\end{array}\biggr|\frac{x}{x-1}\biggr],
 \end{eqnarray*}
 with $\displaystyle
 \lambda=\pi^{-2}\sin\pi\nu\sin\pi\nu'$.
 The kernel $K_{_{\mathrm{G}}}(x,y)$ contains four parameters
 $\nu,\nu',\eta,\eta'\in\mathbb{C}$ chosen so that the Fredholm determinant
 \be\label{2f1det}
 D_{_{\mathrm{G}}}(t)=\mathrm{det}\left(1-K_{_{\mathrm{G}}}|_{(0,t)}\right),\qquad t\in(0,1).
 \eb
 is well-defined. We will not try to determine the set of all possible values of $\nu,\nu',\eta,\eta'$; the interested
 reader may find examples of admissible domains in \cite{bd}.
% Symmetry of the kernel allows to assume that $0\leq\mathrm{Re}\left(\nu+\nu'\right)\leq 1$.

 The above $_2F_1$ kernel first appeared in the harmonic analysis on the infinite-dimensional unitary group
 \cite{bd,olsh}. Later it was shown \cite{dyson2f1} that the determinant (\ref{2f1det}) coincides with
 a correlator of twist fields in the massive Dirac theory on the hyperbolic disk \cite{doyon,lisovyy_JMP,beatty}.
 From the point of view of the present paper, the most interesting feature of $D_{_{\mathrm{G}}}(t)$ is that it is
 a Painlev\'e~VI tau function, see \cite{bd} and also Sec.~5 of \cite{dyson2f1} for a simpler proof. The corresponding
 $P_{_{\mathrm{VI}}}$ parameters are
 \ben
 \left(\theta_0,\theta_t,\theta_1,\theta_{\infty}\right)_{_{\mathrm{VI}}}=\frac12\left(\nu+\nu'+\eta+\eta',0,\nu-\nu',\eta-\eta'\right).
 \ebn
 Monodromy characterizing this particular solution is determined by \cite{dyson2f1}
 \begin{eqnarray*}
 \sigma_{0t}=\frac{\nu+\nu'+\eta+\eta'}{2},\qquad \sigma_{1t}=\frac{\nu+\nu'}{2},\\
 \cos2\pi\sigma_{01}=2e^{-\pi i (\eta+\eta'+\nu+\nu')}\sin\pi\nu\sin\pi\nu'
 +\cos\pi(\eta-\eta').
 \end{eqnarray*}

 Substituting these parameters into (\ref{stpvi}), it can be easily checked that  $\tilde{s}_{_{\mathrm{VI}}}=1$. Remark~\ref{expar1} then implies that the
 large gap ($t\rightarrow1$) expansion of the $_2F_1$ kernel determinant is given by
 \begin{eqnarray}\label{gaussexp1}
  D_{_{\mathrm{G}}}(t)=\chi_{_\mathrm{G}}^{-1}\sum_{n\in\mathbb{Z}}C_{_{\mathrm{G}}}\left(\nu+n,\nu'+n,\eta-n,\eta'-n\right)
 \left(1-t\right)^{(\nu+n)(\nu'+n)}\times\\
 \nonumber \qquad \times\;
 \mathcal{B}_{_{\mathrm{G}}}\left(\nu+n,\nu'+n,\eta-n,\eta'-n;1-t\right) ,
 \end{eqnarray}
 where
 \begin{eqnarray*}
 \fl C_{_{\mathrm{G}}}\left(\nu,\nu',\eta,\eta'\right)=
 G\left[1+\eta,1+\eta',1+\eta+\nu+\nu',1+\eta'+\nu+\nu'\right]\prod_{\epsilon=\pm}G\biggl[\begin{array}{c}1+\epsilon\nu,1+\epsilon\nu' \\
 1+\epsilon\left(\nu+\nu'\right)\end{array}\biggr],\\
 \fl \mathcal{B}_{_{\mathrm{G}}}\left(\nu,\nu',\eta,\eta';1-t\right)=\mathcal{B}_{_{\mathrm{VI}}}\left(\frac{\nu-\nu'}{2},0,
 \frac{\nu+\nu'+\eta+\eta'}{2},\frac{\eta-\eta'}{2},\frac{\nu+\nu'}{2};1-t\right),
 \end{eqnarray*}
 and $\mathcal{B}_{_{\mathrm{VI}}}$ is given by (\ref{cbpvi})--(\ref{bpvi}).
 Also,  \cite[Conjecture~8]{dyson2f1} suggests that the  constant $\chi_{_{\mathrm{G}}}$ is equal to
 \be\label{dysonconj}
 \chi_{_{\mathrm{G}}}=G\Bigl[1+\eta+\nu,1+\eta+\nu',1+\eta'+\nu,1+\eta'+\nu'\Bigr].
 \eb

 Constructing the expansion at $t=0$ is less straightforward. It is of course possible to compute a few first terms
 in the small gap asymptotics directly
 by expanding $D_{_\mathrm{G}}(t)$ into Fredholm series.
 This yields, for instance,
 \ben
  D_{_\mathrm{G}}(t)=1-\kappa_{_\mathrm{G}}t^{1+\eta+\eta'+\nu+\nu'}\left[1+o(1)\right],
 \ebn
 with
 \ben
 \kappa_{_\mathrm{G}}=\lambda\;\Gamma\biggl[\begin{array}{c}1+\eta+\nu,1+\eta'+\nu,1+\eta+\nu',1+\eta'+\nu' \\
 2+\eta+\eta'+\nu+\nu',2+\eta+\eta'+\nu+\nu'\end{array}\biggr].
 \ebn

 On the other hand, direct application of Conjecture~\ref{conjpvi} is ambiguous because of special parameter
 values. First, Barnes functions
 $G\left(1+\theta_t\pm(\theta_0-\sigma_{0t}-n)\right)$ in the structure constants  vanish for $n\gtrless0$. At the same time
 the quantity~$s_{_{\mathrm{VI}}}$ diverges due to zero denominator. The right way to handle this is to fix the values
 of $\theta$'s and~$\sigma_{1t}$, and then consider the limit $\sigma_{0t}\rightarrow\theta_0$ with the help
 of the formulas (\ref{barnesaux1})--(\ref{barnesaux2}) from the Appendix~A. The result is that only
 the terms with $n\geq 0$ survive in the sum over
 $n$ and the structure constants reduce to
 \begin{eqnarray}
 \label{ctgauss}
 \tilde{C}_{_{\mathrm{G}}}(\nu,\nu',\eta,\eta',n)=
 \left(-\lambda\right)^n G\biggl[\begin{array}{c}1+n,1+\eta+\eta'+\nu+\nu'+n \\
 1+\eta+\eta'+\nu+\nu'+2n\end{array}\biggr]^2 \times\\ \nonumber
 \times\; G\biggl[\begin{array}{c}1+\eta+\nu+n,1+\eta'+\nu+n,1+\eta+\nu'+n,1+\eta'+\nu'+n\\
 1+\eta+\nu,1+\eta'+\nu,1+\eta+\nu',1+\eta'+\nu'
 \end{array}\biggr].
 \end{eqnarray}
 In addition, because of the factors $\theta_t+i-j\pm(\sigma_{0t}+n-\theta_0)$ in the products over boxes of $\lambda,\mu\in\mathbb{Y}$
 combinatorial summation in conformal blocks can be restricted to
 Young diagrams  with $\lambda_1\leq n$, $\mu'_1\leq n$.

 This leads to
 the following expansion of $D_{_\mathrm{G}}(t)$ near $t=0$:
 \begin{eqnarray}\label{gaussexp0}
 \fl D_{_{\mathrm{G}}}(t)=\sum_{n=0}^{\infty}\tilde{C}_{_{\mathrm{G}}}\left(\nu,\nu',\eta,\eta',n\right)t^{n(n+\eta+\eta'+\nu+\nu')}
 \!\!\!\sum_{\lambda,\mu\in\mathbb{Y}|\lambda_1,\mu'_1\leq n}\!\!\!
 \mathcal{B}^{^{\mathrm{G}}}_{\lambda,\mu}\left(\nu,\nu',\eta,\eta',n\right)t^{|\lambda|+|\mu|},
 \end{eqnarray}
 where $\tilde{C}_{_{\mathrm{G}}}\left(\nu,\nu',\eta,\eta',n\right)$ is given by (\ref{ctgauss}) and
 \begin{eqnarray*}
 \mathcal{B}^{^{\mathrm{G}}}_{\lambda,\mu}\left(\nu,\nu',\eta,\eta',n\right)= \\ \fl =\!\!\!
 \prod_{(i,j)\in\lambda}
 \frac{\left(i-j+n\right)\left(i-j+n+\eta+\eta'+\nu+\nu'\right)\left(i-j+n+\eta+\nu\right)\left(i-j+n+\eta'+\nu\right)}{
 h_{\lambda}^2(i,j)\left(\lambda'_j+\mu_i-i-j+1+2n+\eta+\eta'+\nu+\nu'\right)^2}\times\\ \fl \times\!\!\!\prod_{(i,j)\in\mu}
 \frac{\left(i-j-n\right)\left(i-j-n-\eta-\eta'-\nu-\nu'\right)\left(i-j-n-\eta-\nu'\right)\left(i-j-n-\eta'-\nu'\right)}{
 h_{\mu}^2(i,j)\left(\lambda_i+\mu'_j-i-j+1-2n-\eta-\eta'-\nu-\nu'\right)^2}\,.
 \end{eqnarray*}
 Note that individual conformal blocks in the sum over $n$ in (\ref{gaussexp0}) give the corresponding terms in the Fredholm series of the
 $_2F_1$~kernel determinant. Numerical checks for randomly chosen $\eta,\eta',\nu,\nu'$ show that the expansions
 (\ref{gaussexp1}) and (\ref{gaussexp0}) perfectly match  for intermediate values of $t$. In particular, this confirms
  the conjectural expression (\ref{dysonconj}).

 \subsubsection{Whittaker kernel.} The Whittaker kernel \cite{whitkernel,bd} emerges in the limit
 \ben
 K_{_{\mathrm{W}}}(x,y)=\lim_{\eta'\rightarrow\infty}\frac{1}{\eta'}\,K_{_{\mathrm{G}}}\Bigl(1-\frac{x}{\eta'},1-\frac{y}{\eta'}\Bigr).
 \ebn
 It contains three parameters $\nu,\nu',\eta$ and has integrable form (\ref{intform}), $\lambda$ is the same as above and
 \begin{eqnarray*}
 \varphi_{_{\mathrm{W}}}(x)=&\,\Gamma\left(1+\eta+\nu\right)\,x^{-\frac12}W_{-\frac{\nu+\nu'+2\eta}{2}+\frac12,\frac{\nu-\nu'}{2}}\left(x\right),\\
 \psi_{_{\mathrm{W}}}(x)=&\,\Gamma\left(1+\eta+\nu'\right)x^{-\frac12}W_{-\frac{\nu+\nu'+2\eta}{2}-\frac12,\frac{\nu-\nu'}{2}}\left(x\right),
 \end{eqnarray*}
 where $W_{k,m}\left(x\right)$ denote the Whittaker functions.

  Fredholm determinant
 \ben
 D_{_{\mathrm{W}}}(t)=\mathrm{det}\left(1-K_{_{\mathrm{W}}}|_{(t,\infty)}\right),\qquad t\in(0,\infty),
 \ebn
 is related to a particular Painlev\'e~V tau function by
 \begin{eqnarray}
 \label{whit1}D_{_{\mathrm{W}}}(t)=t^{-\frac{(\nu-\nu')^2}{4}}\tau_{_{\mathrm{V}}}(t),\\
 \label{whit2}\left(\theta_0,\theta_t,\theta_*\right)_{_{\mathrm{V}}}=\frac12\left(\nu-\nu',0,2\eta+\nu+\nu'\right).
 \end{eqnarray}
 Its expansion around $t=0$ may be found from
 \ben
 D_{_{\mathrm{W}}}(t)=\lim_{\eta'\rightarrow\infty}D_{_{\mathrm{G}}}\Bigl(1-\frac{t}{\eta'}\Bigr).
 \ebn
 Namely, the appropriate termwise limit of (\ref{gaussexp1}) gives
 \be\label{whitexp0}
 \fl D_{_{\mathrm{W}}}(t)=\sum_{n\in\mathbb{Z}}C_{_{\mathrm{W}}}\left(\nu+n,\nu'+n,\eta-n\right)
 t^{(\nu+n)(\nu'+n)}
 \mathcal{B}_{_{\mathrm{W}}}\left(\nu+n,\nu'+n,\eta-n;t\right),
 \eb
 where the limits of structure constants and conformal blocks are
 \begin{eqnarray*}
 & C_{_{\mathrm{W}}}\left(\nu,\nu',\eta\right)=
 G\biggl[\begin{array}{c}1+\nu,1-\nu,1+\nu',1-\nu',1+\eta,1+\eta+\nu+\nu' \\
 1+\nu+\nu',1-\nu-\nu',1+\eta+\nu,1+\eta+\nu'\end{array}\biggr],\\
 & \mathcal{B}_{_{\mathrm{W}}}\left(\nu,\nu',\eta;t\right)=\mathcal{B}_{_{\mathrm{V}}}\left(\frac{\nu-\nu'}{2},0,
 \eta+\frac{\nu+\nu'}{2},\frac{\nu+\nu'}{2};t\right),
 \end{eqnarray*}
 and $\mathcal{B}_{_{\mathrm{V}}}$ was defined in (\ref{CBPV1})--(\ref{CBPV2}).
 Although we are not able to write similar combinatorial expansion  at $t=\infty$, in the latter case
 $D_{_{\mathrm{W}}}(t)$ can still be expanded into Fredholm series. Hence, for example,
 \ben
 \fl D_{_{\mathrm{W}}}(t\rightarrow \infty)=1-\lambda\,\Gamma\left(1+\eta+\nu\right) \Gamma\left(1+\eta+\nu'\right)
 e^{-t}t^{-(2+2\eta+\nu+\nu')}\left[1+O\left(t^{-1}\right)\right].
 \ebn

% Let us mention that  the determinant of Whittaker kernel with $\eta=0$ emerges in the scaling limit of the
% Toeplitz determinant with the symbol (\ref{toep}) as $N\rightarrow\infty$, $t\rightarrow 1$.
 \subsubsection{Confluent hypergeometric kernel.} Another interesting scaling limit of the $_2F_1$ kernel corresponds to setting
 \ben
 \nu'=\nu_0'-i\Lambda,\qquad
 \eta=\eta_0+i\Lambda,
 \ebn
 and then considering
 \ben
 K_{_{\mathrm{F}}}(x,y)=\lim_{\Lambda\rightarrow\infty}\frac{1}{\Lambda}\,K_{_{\mathrm{G}}}\left(\frac{x}{\Lambda},\frac{y}{\Lambda}\right).
 \ebn
 The result is the so-called confluent hypergeometric kernel  \cite{bd,vasilevska}. It depends on three parameters
 \begin{eqnarray*}
 r_+=\nu+\eta',\qquad r_-=\nu_0'+\eta_0,\qquad
 \xi=\frac{1-e^{2\pi i \nu}}{2\pi}\,e^{\frac{i\pi\left(r_--r_+\right)}{2}},
 \end{eqnarray*}
  and is given by (\ref{intform}) with
 \begin{eqnarray*}
 &\lambda=\xi\,\Gamma\Bigl[\begin{array}{cc}
 1+r_+,1+r_- \\ 1+r_++r_-,2+r_++r_-\end{array}\Bigr],\\
 &\varphi_{_{\mathrm{F}}}(x)=x^{1+\frac{r_++r_-}{2}}e^{-\frac{ix}{2}}{}_1F_1\left(r_++1,r_++r_-+2,ix\right),\\
 &\psi_{_{\mathrm{F}}}(x)=x^{\frac{r_++r_-}{2}}e^{-\frac{ix}{2}}{}_1F_1\left(r_+,r_++r_-,ix\right).
 \end{eqnarray*}

 Similarly to (\ref{whit1})--(\ref{whit2}), the $_1F_1$ kernel determinant
 \ben
 D_{_{\mathrm{F}}}(t)=\mathrm{det}\left(1-K_{_{\mathrm{F}}}|_{(0,t)}\right),\qquad t\in(0,\infty),
 \ebn
 can be expressed \cite{bd} in terms of a Painlev\'e~V tau function:
 \begin{eqnarray}
 D_{_{\mathrm{F}}}(t)=t^{-\frac{\left(r_++r_-\right)^2}{4}}\tau_{_{\mathrm{V}}}\left(it\right),\\
 \left(\theta_0,\theta_t,\theta_*\right)_{_{\mathrm{V}}}=\frac12\left(r_++r_-,0,r_+-r_-\right).
 \end{eqnarray}
 Note that
 $ D_{_{\mathrm{F}}}(t)=\lim_{\Lambda\rightarrow\infty}D_{_{\mathrm{G}}}\Bigl(\frac{t}{\Lambda}\Bigr)$. Applying this
 termwise to (\ref{gaussexp0}) and using the properties (\ref{barnesaux0})--(\ref{barnesaux1}) of the Barnes function, we derive the expansion of $D_{_{\mathrm{F}}}(t)$ at $t=0$:
  \begin{eqnarray}\label{1f1exp0}
 \fl D_{_{\mathrm{F}}}(t)=\sum_{n=0}^{\infty}C_{_{\mathrm{F}}}\left(r_+,r_-,n\right)\left(-\xi\right)^n t^{n(n+r_++r_-)}
 \!\!\!\sum_{\lambda,\mu\in\mathbb{Y}|\lambda_1,\mu'_1\leq n}\!\!\!
 \mathcal{B}^{^{\mathrm{F}}}_{\lambda,\mu}\left(r_+,r_-,n\right)\left(it\right)^{|\lambda|+|\mu|},
 \end{eqnarray}
 where
 \begin{eqnarray*}
  C_{_{\mathrm{F}}}\left(r_+,r_-,n\right)=
 G\biggl[\begin{array}{c}1+n,1+r_++r_-+n\\ 1+r_++r_-+2n\end{array}\biggr]^2 G\biggl[\begin{array}{c}1+r_++n,1+r_-+n \\
 1+r_+,1+r_-\end{array}\biggr],\\
   \mathcal{B}^{^{\mathrm{F}}}_{\lambda,\mu}\left(r_+,r_-,n\right)=
 \prod_{(i,j)\in\lambda}
 \frac{\left(i-j+n\right)\left(i-j+n+r_+\right)\left(i-j+n+r_++r_-\right)}{
 h_{\lambda}^2(i,j)\left(\lambda'_j+\mu_i-i-j+1+2n+r_++r_-\right)^2}\times\\ \times\prod_{(i,j)\in\mu}
 \frac{\left(i-j-n\right)\left(i-j-n-r_-\right)\left(i-j-n-r_+-r_-\right)}{
 h_{\mu}^2(i,j)\left(\lambda_i+\mu'_j-i-j+1-2n-r_+-r_-\right)^2}\,.
 \end{eqnarray*}

 \subsubsection{Sine kernel.}
  Certain specializations of the $_1F_1$ kernel play an important role in
  random matrix theory. In particular, for $r_+=r_-=r$ it coincides with the  Bessel kernel \cite{Nagao,FWBessel}
  \ben
   K_{_{\mathrm{B}}}(x,y)=\frac{\pi\xi\sqrt{xy}}{2}\frac{J_{r+\frac12}\left(\frac{x}{2}\right)J_{r-\frac12}\left(\frac{y}{2}\right)-
   J_{r-\frac12}\left(\frac{x}{2}\right)J_{r+\frac12}\left(\frac{y}{2}\right)}{x-y},
  \ebn
  which in the case $r=0$ reduces to the celebrated
 sine kernel
 \ben
 K_{\mathrm{sine}}(x,y)=\frac{2\xi\sin\frac{x-y}{2}}{x-y}.
 \ebn

 It is well-known that the determinant
 \ben
 D_{\mathrm{sine}}\left(t\right)=\mathrm{det}\left(1-K_{\mathrm{sine}}|_{(0,t)}\right)
 \ebn
 for $\xi=\frac{1}{2\pi}$ coincides with the scaled gap probability in the bulk of the Gaussian Unitary Ensemble
 \cite{Forrester_book}. The expansion (\ref{1f1exp0}) thus gives a complete series for this quantity:
 \be\label{gueexp}
 D_{\mathrm{sine}}\left(t\right)=\sum_{n=0}^{\infty}\frac{G^6(1+n)}{G^2(1+2n)}\left(-\xi\right)^n
 t^{n^2}\!\!\! \!\!\!\sum_{\lambda,\mu\in\mathbb{Y}|\lambda_1,\mu'_1\leq n}\!\!\!
 \mathcal{B}^{^{\mathrm{sine}}}_{\lambda,\mu}\left(n\right)\left(it\right)^{|\lambda|+|\mu|},
 \eb
 where
 \begin{eqnarray*}
  \mathcal{B}^{^{\mathrm{sine}}}_{\lambda,\mu}\left(n\right)&=
 \prod_{(i,j)\in\lambda}
 \frac{\left(i-j+n\right)^3}{
 h_{\lambda}^2(i,j)\left(\lambda'_j+\mu_i-i-j+1+2n\right)^2}\times \\ &\times\prod_{(i,j)\in\mu}
 \frac{\left(i-j-n\right)^3}{
 h_{\mu}^2(i,j)\left(\lambda_i+\mu'_j-i-j+1-2n\right)^2}\,.
 \end{eqnarray*}

 First terms of the series (\ref{gueexp}) are recorded in the Appendix~B. In particular,
 they reproduce the results obtained by an iterative expansion of the corresponding
 Painlev\'e~V solution, cf Eq.~(8.114) in~\cite{Forrester_book}. Note that our $t=2\pi t_{\tiny\cite{Forrester_book}}$, $\xi=\frac{\xi_{\tiny\cite{Forrester_book}}}{2\pi}$. We have also checked the agreement of (\ref{gueexp}) with the
  known large gap ($t\rightarrow\infty$) asymptotics \cite{dyson}
 \ben
 D_{\mathrm{sine}}\left(4t\right)\Bigr|_{\xi=\frac{1}{2\pi}}=\sqrt{\pi}\,G^2\left(\frac12\right)t^{-\frac14}
 e^{-\frac{t^2}{2}}\left[1+\frac{1}{32}\,t^{-2}+\frac{81}{2048}t^{-4}+O\left(t^{-6}\right)\right].
 \ebn

% \begin{comment}
 \subsubsection{Modified Bessel kernel.} One may also study a further scaling limit of the $_1F_1$ kernel by setting
 \ben
 r_{\pm}=\frac{r}{2}\mp i\Lambda,\qquad \xi=\xi_{_{\mathrm{B2}}}\frac{re^{\pi\Lambda}}{2\pi},
 \ebn
 and defining
 \ben
 K_{_{\mathrm{B2}}}(x,y)=\lim_{\Lambda\rightarrow\infty}\frac{1}{\Lambda}K_{_{\mathrm{F}}}\left(\frac{x}{\Lambda},\frac{y}{\Lambda}\right).
 \ebn
  Asymptotic properties of the confluent hypergeometric function imply that
 \ben
 K_{_{\mathrm{B2}}}(x,y)=\xi_{_{\mathrm{B2}}}\sqrt{xy}\,\frac{I_{r+1}\left(2\sqrt{x}\right)I_{r-1}\left(2\sqrt{y}\right)-
 I_{r-1}\left(2\sqrt{x}\right)I_{r+1}\left(2\sqrt{y}\right)}{x-y}.
 \ebn

 Fredholm determinant $D_{_{\mathrm{B2}}}(t)=\mathrm{det}\left(1-K_{_{\mathrm{B2}}}|_{(0,t)}\right)$ is related to a  tau function
 of Painlev\'e~$\mathrm{III}'_1$ with $\theta_*=-\theta_{\star}=\frac{r}{2}$ by
 \be\label{bessel2}
 D_{_{\mathrm{B2}}}(t)=t^{-\frac{r^2}{4}}e^{\frac{t}{2}}\tau_{_{\mathrm{III}'_1}}(t).
 \eb
 Its small gap expansion can be
 calculated using that $D_{_{\mathrm{B2}}}(t)=\lim_{\Lambda\rightarrow\infty}D_{_{\mathrm{F}}}\Bigl(\frac{t}{\Lambda}\Bigr)$. We find
 \begin{eqnarray}
 \label{besselexp}\fl D_{_{\mathrm{B2}}}(t)=\sum_{n=0}^{\infty}G\biggl[\begin{array}{c}1+n,1+r+n\\ 1+r+2n\end{array}\biggr]^2\left(-\xi_{_{\mathrm{B2}}}r\right)^nt^{n(n+r)}
 \!\!\! \!\!\!\sum_{\lambda,\mu\in\mathbb{Y}|\lambda_1,\mu'_1\leq n}\!\!\!
 \mathcal{B}^{^{\mathrm{B2}}}_{\lambda,\mu}\left(r,n\right)t^{|\lambda|+|\mu|},
 \end{eqnarray}
 \begin{eqnarray*}
 \nonumber  \mathcal{B}^{^{\mathrm{B2}}}_{\lambda,\mu}\left(r,n\right)&=
 \prod_{(i,j)\in\lambda}
 \frac{\left(i-j+n\right)\left(i-j+n+r\right)}{
 h_{\lambda}^2(i,j)\left(\lambda'_j+\mu_i-i-j+1+2n+r\right)^2}\times\\
 \nonumber  &\times\,\prod_{(i,j)\in\mu}
 \frac{\left(i-j-n\right)\left(i-j-n-r\right)}{
 h_{\mu}^2(i,j)\left(\lambda_i+\mu'_j-i-j+1-2n-r\right)^2}\,.
 \end{eqnarray*}
% \end{comment}

 \subsection{Sine-Gordon exponential fields}
 A well-known example of appearance of Painlev\'e transcendents
 in integrable QFT is provided by the
 two-point correlation function of exponential fields
 $Q\left(mr\right)=\left\langle\mathcal{O}_{\nu}\left(0\right)\mathcal{O}_{\nu'}(r)\right\rangle$
 in the sine-Gordon model at the free-fermion point \cite{BL,smj_IV}. The spectrum of this model consists of
 fermionic excitations of mass $m$, parameterized by the topological charge $\epsilon=\pm1$ and rapidity $\theta\in\mathbb{R}$.
 Lattice counterparts of the exponential fields have been introduced and studied in \cite{GIL,Palmer}.
 \subsubsection{From form factors to Macdonald kernel.}
 Under normalization $\langle \mathcal{O}_{\nu}\rangle=1$, the exponential fields are completely determined by their two-particle form factors~\cite{schroer}
 \ben
 \mathcal{F}_{\nu}\left(\theta,\theta'\right)={}^{+-}\!\left\langle \theta;\theta' |\mathcal{O}_{\nu}(0)|vac\right\rangle=
 \frac{i\sin\pi\nu}{2\pi}
 \frac{ e^{\nu\left(\theta'-\theta\right)}}{\cosh\frac{\theta'-\theta}{2}}.
 \ebn
 Multiparticle form factors can be written as determinants of two-particle ones. This allows
 to sum up the form factor expansion
 \begin{eqnarray*}
 Q\left(mr\right)=&\,\sum_{n=0}^{\infty}\sum_{\epsilon_1,\ldots,\epsilon_n=\pm}
 \frac{1}{n!}\int_{-\infty}^{\infty}
 \ldots\int_{-\infty}^{\infty} d\theta_1
 \ldots d\theta_n\,e^{-mr\sum_{k=1}^{n}\cosh\theta_k}\times\\
 &\,\times{}\left\langle vac|\mathcal{O}_{\nu}(0)|\theta_1,\ldots,\theta_n\right\rangle_{\epsilon_1,\ldots,\epsilon_n}
 \!\!\!\!\!\!{}^{\epsilon_1,\ldots,\epsilon_n}\left\langle\theta_1,\ldots,\theta_n|\mathcal{O}_{\nu'}(0)|vac\right\rangle
 \end{eqnarray*}
 to Fredholm determinant $Q\left(mr\right)=\mathrm{det}\left(1-K_{_{\mathrm{SG}}}\right)$. The corresponding kernel
 acts on $L^2(\mathbb{R})$ and is expressed in terms
 of dressed two-particle form factors:
 \ben
 K_{_{\mathrm{SG}}}(\theta,\theta')=\int_{-\infty}^{\infty}\mathcal{F}_{-\nu}\left(\theta'',\theta\right)\mathcal{F}_{\nu'}\left(\theta'',\theta'\right)
 e^{-\frac{mr}{2}\left(\cosh\theta+2\cosh\theta''+\cosh\theta'\right)}d\theta''.
 \ebn

 Let us show that  $K_{_{\mathrm{SG}}}(\theta,\theta')$
 is equivalent to a more familiar classical integrable kernel $K_{_{\mathrm{M}}}(x,y)$ on $L^2\bigl(\frac{m^2r^2}{4},\infty\bigr)$. The latter is
  defined by (\ref{intform}) with $\lambda=\pi^{-2}\sin\pi\nu\sin\pi\nu'$ and $\varphi$, $\psi$ given by Macdonald functions
 \be\label{mcd2}
 \varphi_{_{\mathrm{M}}}(x)=2\sqrt{x}K_{\nu'-\nu+1}\left(2\sqrt{x}\right),\qquad \psi_{_{\mathrm{M}}}(x)=2K_{\nu'-\nu}\left(2\sqrt{x}\right).
 \eb
 This kernel can be seen as a further scaling limit of the Whittaker kernel from the previous subsection. Indeed,
 one may check that
 \ben
 K_{_{\mathrm{M}}}(x,y)=\lim_{\eta\rightarrow\infty}\frac{1}{\eta}\,K_{_{\mathrm{W}}}\left(\frac{x}{\eta},\frac{y}{\eta}\right).
 \ebn
 By equivalence of $K_{_{\mathrm{SG}}}$ and $K_{_{\mathrm{M}}}$ we mean that $\mathrm{Tr}\, K_{_{\mathrm{SG}}}^n=\mathrm{Tr}\,K_{_{\mathrm{M}}}^n$ for any $n\in\mathbb{Z}_{\geq 0}$.

 Here is a proof. First note that the Macdonald kernel admits an alternative simple form
 \be\label{mcd3}
 K_{_{\mathrm{M}}}(x,y)=\lambda\int_1^{\infty} \psi_{_{\mathrm{M}}}(xt)\psi_{_{\mathrm{M}}}(yt)\,dt.
 \eb
 This representation results from the identity
 \ben
 \frac{d}{dt}\left[\varphi_{_{\mathrm{M}}}(xt)\psi_{_{\mathrm{M}}}(yt)-\varphi_{_{\mathrm{M}}}(yt)\psi_{_{\mathrm{M}}}(xt)\right]=
 -(x-y)\,\psi_{_{\mathrm{M}}}(xt)\psi_{_{\mathrm{M}}}(yt),
 \ebn
 which is itself an easy consequence of the differentiation formulas
 \ben
 x\frac{d}{dx}\left(\begin{array}{c}\varphi_{_{\mathrm{M}}}(x) \\ \psi_{_{\mathrm{M}}}(x)\end{array}\right)=
 \left(\begin{array}{cc} \frac{\nu-\nu'}{2} & -x \\ -1 & \frac{\nu'-\nu}{2}\end{array}\right)
 \left(\begin{array}{c}\varphi_{_{\mathrm{M}}}(x) \\ \psi_{_{\mathrm{M}}}(x)\end{array}\right).
 \ebn
 On the other hand, parameterizing the rapidities as $u=e^{\theta}$, one can write $\kappa_n=\mathrm{Tr}\,K_{_{\mathrm{SG}}}^n$ as
 \be\label{kappan}
 \kappa_n=\lambda^n\int_0^{\infty}\ldots\int_0^{\infty}du_1\ldots du_{2n}\prod_{j=1}^n\frac{u_{2j-1}^{\nu'-\nu}}{u_{2j}^{\nu'-\nu}}
 \prod_{j=1}^{2n}\frac{\exp\left\{-\frac{mr}{2}\left(u_j+u_j^{-1}\right)\right\}}{u_{j}+u_{j+1}},
 \eb
 with $u_{2n+1}=u_1$. Now make in (\ref{kappan}) the following replacements:
 \begin{eqnarray*}
 & \frac{e^{-\frac{mr}{2}\left(u_{2j-1}+u_{2j}\right)}}{u_{2j-1}+u_{2j}}=\int_{\frac{mr}{2}}^{\infty}
 e^{-t_{2j-1}\left(u_{2j-1}+u_{2j}\right)}dt_{2j-1},\\
 & \frac{e^{-\frac{mr}{2}\left(u_{2j}^{-1}+u_{2j+1}^{-1}\right)}}{u_{2j-1}+u_{2j}}=u_{2j}^{-1}u_{2j+1}^{-1}
 \int_{\frac{mr}{2}}^{\infty}
 e^{-t_{2j}\left(u_{2j}^{-1}+u_{2j+1}^{-1}\right)}dt_{2j-1},
 \end{eqnarray*}
 where $j=1,\ldots,n$. This yields a $4n$-fold integral
 \ben
 \fl \int_0^{\infty}\frac{du_1}{u_1}\ldots\int_0^{\infty}\frac{du_{2n}}{u_{2n}}
 \int_{\frac{mr}{2}}^{\infty}dt_1\ldots\int_{\frac{mr}{2}}^{\infty}dt_{2n}
 \prod_{j=1}^n\frac{u_{2j-1}^{\nu'-\nu}}{u_{2j}^{\nu'-\nu}}\,e^{-t_{2j-1}u_{2j-1}-t_{2j-2}u_{2j-1}^{-1}-t_{2j-1}u_{2j}-t_{2j}u_{2j}^{-1}},
 \ebn
 with $t_0=t_{2n}$. The variables $u_1,\ldots,u_{2n}$ are now decoupled. Integrating them out with the help of the standard
 integral representation of the Macdonald function
 \ben
 \int_{0}^{\infty}u^{-1\pm(\nu'-\nu)}e^{-tu-t'u^{-1}}du=\left(t/t'\right)^{\mp\frac{\nu'-\nu}{2}}\psi_{_{\mathrm{M}}}(tt'),\qquad t,t'>0,
 \ebn
 we finally obtain
 \ben
 \kappa_n=\lambda^n\int_{\frac{mr}{2}}^{\infty}dt_1\ldots\int_{\frac{mr}{2}}^{\infty}dt_{2n}\prod_{j=1}^{2n}\psi_{_{\mathrm{M}}}(t_{j-1}t_j).
 \ebn
 After the change of variables $t_{2j-1}\mapsto \frac{mr}{2}t_{2j-1}$, $t_{2j}\mapsto \frac{2}{mr}t_{2j}$ the last expression
 can obviously be written as $\mathrm{Tr}\,K_{_{\mathrm{M}}}^n$ with $K_{_{\mathrm{M}}}$ given by (\ref{mcd3}).

 \subsubsection{Painlev\'e III and asymptotics.}
 Painlev\'e representations of the two-point function of exponential fields \cite{BL,smj_IV} can now be rederived by applying the
 standard random matrix theory techniques \cite{tweqs} to the Macdonald kernel.
 The final result is that $Q(mr)=\mathrm{det}\left(1-K_{_{\mathrm{M}}}|_{\bigl(\frac{m^2r^2}{4},\infty\bigr)}\right)$ coincides, up to a simple prefactor, with a tau function of Painlev\'e~$\mathrm{III}'_1$ equation
  with parameters $\theta_*=-\theta_{\star}=\frac{\nu-\nu'}{2}$:
 \ben
 Q\bigl(2\sqrt{t}\bigr)=t^{-\frac{(\nu-\nu')^2}{4}}e^{\frac{t}{2}}
 \tau_{_{\mathrm{III}'_1}}\left(t\right).
 \ebn
 The integration constants specifying this tau function are \cite{jimbo}
 \be\label{sgconst}
 \sigma=\frac{\nu+\nu'}{2}, \qquad s_{_{\mathrm{III}'_1}}=1.
 \eb

 In general, the tau function is defined up to multiplication by a constant, which in the case at hand is fixed by
 normalization of the VEVs: $Q(mr)\simeq 1$ as $r\rightarrow\infty$. Subleading corrections to this long-distance
 behaviour can be obtained from the form factor expansion. For instance, taking into account the  contribution
 of two-particle states,  we find
 \ben
 1-Q\bigl(2\sqrt{t}\bigr)=\lambda \underbrace{\int_{t}^{\infty}\Bigl(\varphi_{_{\mathrm{M}}}'(x)\psi_{_{\mathrm{M}}}(x)-
 \varphi_{_{\mathrm{M}}}(x)\psi_{_{\mathrm{M}}}'(x)\Bigr)\,dx}_{O\left(t^{-1/2}e^{-4\sqrt{t}}\right)}+
 O\left(t^{-1}e^{-8\sqrt{t}}\right).
 \ebn
 Short-distance asymptotics of $Q(mr)$ is also known. Assume that $|\mathrm{Re}\left(\nu+\nu'\right)|<1$, then, as $t\rightarrow0$,
 \ben
 Q\bigl(2\sqrt{t}\bigr)\simeq C_{_{\mathrm{SG}}}\left(\nu,\nu'\right)t^{\nu\nu'}.
 \ebn
 The value of $\sigma$ in (\ref{sgconst}) is determined by the exponent $\nu\nu'$, found in \cite{jimbo}.
 The coefficient $C_{_{\mathrm{SG}}}\left(\nu,\nu'\right)$ was calculated by Basor and Tracy in \cite{basor}:
 \begin{eqnarray}\label{csg}
 C_{_{\mathrm{SG}}}\left(\nu,\nu'\right)=G\biggl[\begin{array}{c}1+\nu,1-\nu,1+\nu',1-\nu' \\
 1+\nu+\nu',1-\nu-\nu'\end{array}\biggr].
 \end{eqnarray}
 Note that the last expression coincides with $C_{_{\mathrm{III}'_1}}\Bigl(\frac{\nu-\nu'}{2},\frac{\nu'-\nu}{2},\frac{\nu+\nu'}{2}\Bigr)$
 defined by (\ref{cp3v1}). This simply means that the normalization of $\tau_{_{\mathrm{III}'_1}}(t)$ in Conjecture~\ref{conjp31}
 corresponds to setting $\langle \mathcal{O}_{\nu}\rangle=1$ in the sine-Gordon case.

 \subsubsection{Short-distance expansion of $\left\langle\mathcal{O}_{\nu}\left(0\right)\mathcal{O}_{\nu'}(r)\right\rangle$.}
 We are now ready to write complete short-distance expansion of the two-point correlator $Q(mr)$.
 Combining the above with Conjecture~\ref{conjp31} gives the following series:
 \begin{eqnarray}\label{qexp}
 \fl Q(mr)= \sum_{n\in\mathbb{Z}}C_{_{\mathrm{SG}}}(\nu+n,\nu'+n)
 \sum_{\lambda,\mu\in\mathbb{Y}}
 \mathcal{B}^{^{\mathrm{SG}}}_{\lambda,\mu}\left(\nu+n,\nu'+n\right)\left(\frac{m^2r^2}{4}\right)^{(\nu+n)(\nu'+n)+|\lambda|+|\mu|},
 \end{eqnarray}
 where $C_{_{\mathrm{SG}}}(\nu,\nu')$ is defined by (\ref{csg}) and
  \begin{eqnarray*}
 \mathcal{B}^{^{\mathrm{SG}}}_{\lambda,\mu}\left(\nu,\nu'\right)&=
 \prod_{(i,j)\in\lambda}
 \frac{\left(i-j+\nu\right)\left(i-j+\nu'\right)}{
 h_{\lambda}^2(i,j)\left(\lambda'_j+\mu_i-i-j+1+\nu+\nu'\right)^2}\times\\ &\times\,\prod_{(i,j)\in\mu}
 \frac{\left(i-j-\nu\right)\left(i-j-\nu'\right)}{
 h_{\mu}^2(i,j)\left(\lambda_i+\mu'_j-i-j+1-\nu-\nu'\right)^2}\,.
 \end{eqnarray*}
 The series (\ref{qexp}) has a familiar structure of conformal perturbation expansion \cite{FFLZZ,Zamo_LY}. The non-analytic
 factors $m^{2(\nu+n)(\nu'+n)}$ correspond to non-perturbative VEVs of the primary fields which appear in the operator
 product expansion $\mathcal{O}_{\nu}\left(0\right)\mathcal{O}_{\nu'}(r)$. All other corrections, including the VEVs of descendant fields and CPT, come in integer powers
 of the coupling~$m^2$.

 Fig.~5 illustrates how well
 the series (\ref{qexp}) fits form factor expansion to give all-distance behaviour of the correlator.
 We fix $\nu=0.3$, $\nu'=0.45$ and compute the expansion $Q_{_{15}}(mr)$ taking into account
 the terms with $n=-4,\ldots,4$ up to descendant
 level~15, as we did before for $P_{_{\mathrm{VI}}}$. Plots (A), (B), (C), (D) correspond to the logarithms of $-\sum_{j=1}^{\ell-1}\frac{1}{j}\mathrm{Tr}\,K^{j}_{_{\mathrm{SG}}}-\ln Q_{_{15}}(mr)$ (solid lines) and $\frac{1}{\ell}\mathrm{Tr}\,K^{\ell}_{_{\mathrm{SG}}}$ (dotted lines) for $\ell=1,2,3,4$.
 Hence $Q_{_{15}}(mr)$ correctly accounts for the 2-particle form factor contribution to long-distance asymptotics
 up to $mr\approx 3.2$, 4-particle contribution up to $mr \approx 2.6$, 6-particle and 8-particle ones up to $mr \approx 2.1$ and $mr\approx 1.7$.

  \begin{figure}[!h]
 \begin{center}
 \resizebox{10cm}{!}{
 \includegraphics{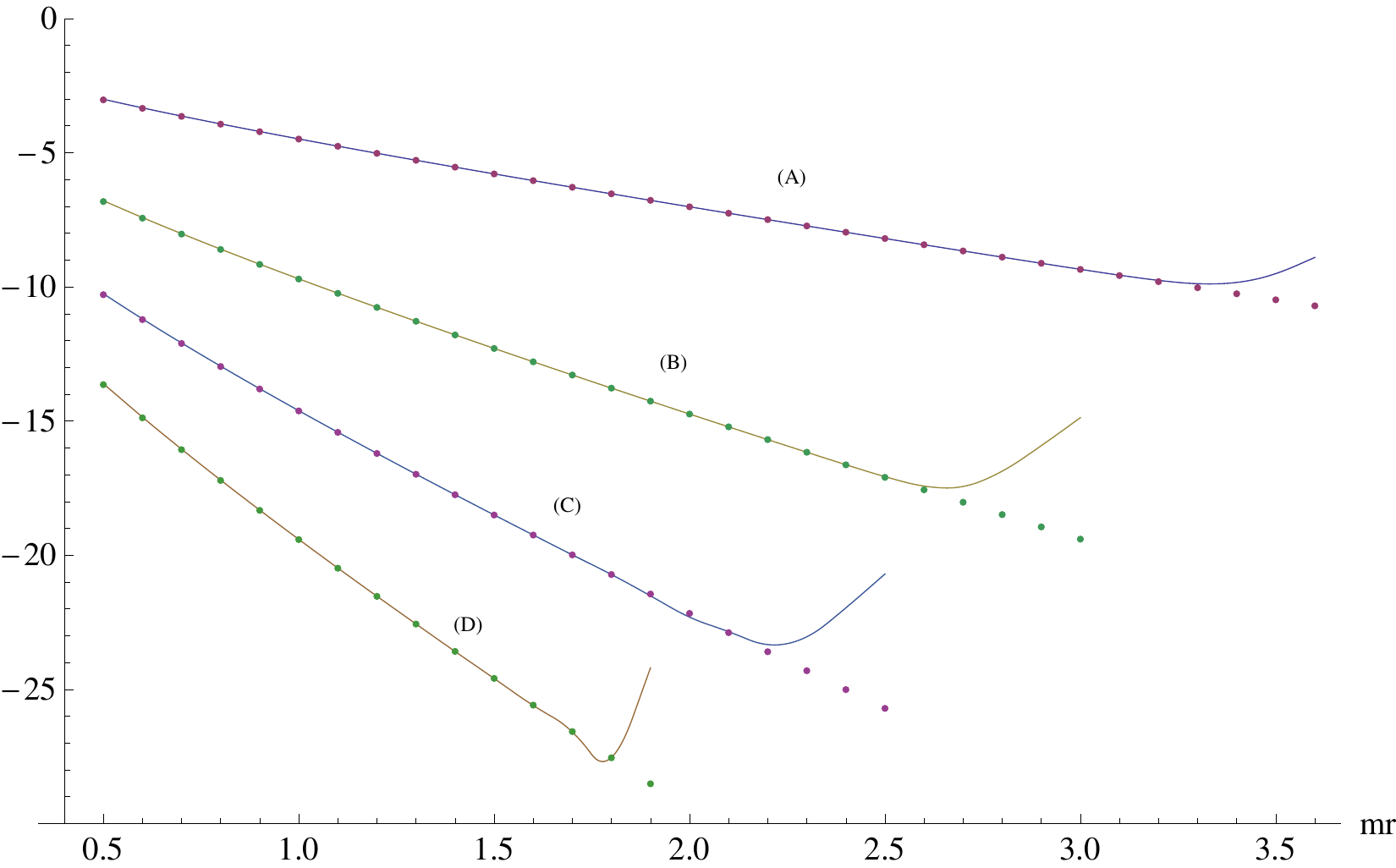}} \\
 Fig. 5: Short-distance vs form factor expansion
 \end{center}
 \end{figure}

 Let us stress that we are dealing here with a correlation function of \textit{massive} theory. Its description
 by \textit{holomorphic}  conformal blocks  therefore looks rather surprising and is presumably related to
 the affine $\widehat{sl(2)}$ symmetry of the free-fermion sine-Gordon theory \cite{leclair}.
 \subsubsection{Painlev\'e III and 2D polymers.}
 The change of variables $q_{_{\mathrm{III}'_1}}(t)=\frac{r}{4}\exp\psi(r)$, $r=4\sqrt{t}$ maps
 $P_{_{\mathrm{III}'_1}}$ with $\theta_{\star}=0$, $\theta_{*}=\frac12$ to radial
 sinh-Gordon equation
 \be\label{rsg}
 \psi''+\frac1r \psi'=\frac12\sinh2\psi.
 \eb
 A particular solution of this equation describes universal scaling functions of 2D polymers \cite{FS,Zamo_PIII}. It is
 characterized by the boundary conditions
 \be\label{bcsg}
 \psi(r\rightarrow0)\sim -\frac13\ln r-
 \frac12\ln\frac{\mu}{4}+O\left(r^{4/3}\right),\qquad\mu=\frac{\Gamma^2(1/3)}{\Gamma^2(2/3)},
 \eb
 which in our notation correspond to integration constants $\sigma_{_{\mathrm{III}'_1}}=\frac16$, $s_{_{\mathrm{III}'_1}}=1$.
 In fact the relevant $P_{_{\mathrm{III}'_1}}$  solution is a B\"acklund transform of a solution associated
 to the tau function considered in the previous subsection. The
 precise relation between the two quantities is
 \ben
 \sinh^2\psi(r)=-\Bigl[\left(\ln Q(r)\right)''+r^{-1}\left(\ln Q(r)\right)'\Bigr]_{\nu=\nu'=\frac16}.
 \ebn

  On the other hand, it is known \cite{ohyama} that $P_{_{\mathrm{III_1'}}}$
  with $\theta_{\star}=0$, $\theta_*=\frac12$ (and hence the radial sinh-Gordon equation!) is equivalent to $P_{_{\mathrm{III_3'}}}$. Namely,
 if we set
 \ben
 t_{_{\mathrm{III_3'}}}=\frac{t^2}{16},\qquad q_{_{\mathrm{III_3'}}}\left(t_{_{\mathrm{III_3'}}}\right)=\frac{q^2(t)}{4},
 \ebn
 then $q(t)$ satisfies appropriate $P_{_{\mathrm{III_1'}}}$. This allows to give an alternative characterization of the
 solution (\ref{bcsg}) via the expansion (\ref{piii3first})--(\ref{piii3last}):
 \ben
 e^{-2\psi(r)}=-4r^{-1} \frac{d}{dr} r\frac{d}{dr}\ln \tau_{_{\mathrm{III_3'}}}\left(2^{-12}r^4\right)\Bigl|_{
 s_{_{\mathrm{III_3'}}}=1,\sigma=\frac16},
 \ebn
 or, in yet another form,
 \ben
 e^{\psi(r)}=\frac{\tau_{_{\mathrm{III_3'}}}\left(2^{-12}r^4\right)\Bigl|_{
 s_{_{\mathrm{III_3'}}}=1,\sigma=\frac16}}{\tau_{_{\mathrm{III_3'}}}\left(2^{-12}r^4\right)\Bigl|_{
 s_{_{\mathrm{III_3'}}}=1,\sigma=\frac13}},
 \ebn
 where the normalization of both tau functions in the last formula is precisely the same as in Conjecture~\ref{conjpiii3}.

 \section{Discussion}
 We believe that by explaining the title of this paper we have partially answered P. Deift's question from the Introduction.
 Besides the obvious need for rigorous proofs of our claims in Section~\ref{secsols}, many other questions beg to be addressed.
 Why instantons? Is there
 a way to obtain irregular ``form factor'' expansions at $\infty$ for general solutions of $P_{_{\mathrm{V}}}$ and $P_{_{\mathrm{III}}}$?
 What about $P_{_{\mathrm{IV}}}$, $P_{_{\mathrm{II}}}$ and $P_{_{\mathrm{I}}}$?

  A particularly interesting problem,
 already mentioned above, concerns the computation of connection coefficients of Painlev\'e tau functions
 (akin to Dyson-Widom constants in random matrix theory). In the $P_{_{\mathrm{VI}}}$
 case this is very much related to determining the fusion matrix for $c=1$ generic conformal blocks.

 Another intriguing issue is the quantization of Painlev\'e equations \cite{nagoya,zabrodin}.
 The existing paradigm usually associates isomonodromic deformations to semiclassical limit of CFT
 \cite{teschner}. For instance, the scalar Lax pairs for $P_{_{\mathrm{VI-I}}}$ emerge in the
 $c\rightarrow \infty$ limit of two BPZ-type differential operators \cite{yamada}. The results presented here and in \cite{CFT_PVI}
 suggest a completely different, $c=1$ point of view. In this picture, classical Riccati solutions of Painlev\'e equations
 may be naturally deformed to Coulomb $\beta$-integrals. It would be nice to
 understand whether the general case allows for a similar $\beta$-deformation.

 \ack
 We are grateful to P. Gavrylenko and V. Shadura for useful discussions.
 The present work  was supported by the  ERC grant 279738-NEDFOQ (O.~Gamayun),
 the Joint Ukrainian-Russian SFFR-RFBR project F53.2 and
  the Program of fundamental research of the physics and astronomy division of NASU  (N.~Iorgov),
  the IRSES project ``Random and integrable models in mathematical physics''
 (O.~Lisovyy),  and the joint program of bilateral seminars of CNRS and NASU.

 \appendix
 \section*{Appendix A. Barnes function}
 \setcounter{section}{1}
% \subsection{Barnes function}
 Barnes $G$-function satisfies the functional equation $G\left(1+z\right)=\Gamma\left(z\right)G\left(z\right)$
 and is defined as the infinite product
 \ben
 G\left(1+z\right)=\left(2\pi\right)^{\frac{z}{2}}\exp\left(-\frac{z+z^2\left(1+\gamma\right)}{2}\right)\prod_{k=1}^{\infty}
 \left(1+\frac{z}{k}\right)^k \exp\left(\frac{z^2}{2k}-z\right),
 \ebn
 where $\gamma$ is the Euler's constant, or via the integral representation
 \ben
 G(1+z)=\left(2\pi\right)^{\frac{z}{2}}\exp\int_0^{\infty}\frac{dt}{t}\Biggl[\frac{1-e^{-zt}}{4\sinh^2\frac{t}{2}}-\frac{z}{t}+
 \frac{z^2}{2}\,e^{-t}\Biggr],\qquad \mathrm{Re}\,z>-1.
 \ebn
   It is analytic in the whole complex plane and has
   the following asymptotic expansion as $|z|\rightarrow\infty$, $\mathrm{arg}\,z\neq\pi$:
  \ben
  \ln G(1+z)=\left(\frac{z^2}{2}-\frac{1}{12}\right)\ln z-\frac{3z^2}{4}
  +\frac{z}{2}\ln 2\pi+\zeta'(-1)+O\left(z^{-2}\right).
  \ebn
   One of the consequences
  of this asymptotic  behaviour is the formula
  \be\label{barnesaux0}
  G\biggl[\begin{array}{c}1+z+\alpha,1+z-\alpha\\ 1+z+\beta,1+z-\beta\end{array}\biggr]=
  z^{\alpha^2-\beta^2}\left[1+O\left(z^{-2}\right)\right].
  \eb
 Another useful relation is
 \be\label{barnesaux1}
 G\biggl[\begin{array}{c}1+z+n,1-z\\ 1-z-n,1+z\end{array}\biggr]=\left(-1\right)^{\frac{n(n+1)}{2}}\left(\frac{\pi}{\sin\pi z}\right)^n,\qquad n\in\mathbb{Z}.
 \eb
  It is easy to deduce from it that, as $\varepsilon\rightarrow0$,
 \be\label{barnesaux2}
 G\left(1+\varepsilon-n\right)\sim \varepsilon^n\left(-1\right)^{\frac{n(n-1)}{2}}G\left(1+n\right),\qquad n\in\mathbb{Z}_{\geq0}.
 \eb
  \appendix
 \section*{Appendix B. Sine kernel conformal blocks}
 \setcounter{section}{1}
% \subsection{Sine kernel conformal blocks}
 Consider the functions
 \ben
 \mathcal{B}_{_{\mathrm{sine}}}(n;t)=\!\!\!\!\!\!\sum_{\lambda,\mu\in\mathbb{Y}|\lambda_1,\mu'_1\leq n}\!\!\!
 \mathcal{B}^{^{\mathrm{sine}}}_{\lambda,\mu}\left(n\right)\left(it\right)^{|\lambda|+|\mu|},
 \ebn
 which appear in the expansion (\ref{gueexp}) of the GUE gap probability.  Below we record the terms contributing
 to $D_{\mathrm{sine}}\left(t\right)$ as at least $t^{30}$:
 \begin{eqnarray*}
 \fl \mathcal{B}_{_{\mathrm{sine}}}(0;t)=\mathcal{B}_{_{\mathrm{sine}}}(1;t)=1,\\
 \fl\mathcal{B}_{_{\mathrm{sine}}}(2;t)=1-\frac{t^2}{75}+\frac{t^4}{7840}-\frac{t^6}{1134000}+\frac{t^8}{219542400}-
 \frac{t^{10}}{55091836800}+\frac{t^{12}}{17435658240000}\\
 \fl\qquad  -\frac{t^{14}}{6802522062336000}+\frac{t^{16}}{3210079038566400000}-\frac{t^{18}}{1803084500809912320000}\,+\\
 \fl\qquad  +\frac{t^{20}}{1189192769988708925440000}-\frac{t^{22}}{910206422681575219200000000}+\\
 \fl\qquad  +\frac{t^{24}}{800331904605748883816448000000}-\frac{t^{26}}{801284680682660489630515200000000}+O\left(t^{28}\right),\\
 \fl \mathcal{B}_{_{\mathrm{sine}}}(3;t)=1-\frac{18\,t^2}{1225}+\frac{t^4}{8820}-\frac{2293\,t^6}{3922033500}+
 \frac{3581\,t^8}{1616027212800}-\frac{71\,t^{10}}{10908183686400}+\\
 \fl\qquad  +\frac{94789\,t^{12}}{6178831567324416000}-\frac{76477\,t^{14}}{2570452778021883955200}\\
 \fl\qquad  +\frac{407221\,t^{16}}{8412390909889802035200000}
  -\frac{245265109\,t^{18}}{3655090136312382811899727872000}\\
  \fl\qquad +\frac{40956413\,t^{20}}{510254748374093327017340928000000}
 +O\left(t^{22}\right),\\
 \fl \mathcal{B}_{_{\mathrm{sine}}}(4;t)=1-\frac{20\,t^2}{1323}+\frac{83\,t^4}{711480}-\frac{174931\,t^6}{286339821768}+
 \frac{9605\,t^8}{3926946127104}\\
 \fl\qquad -\frac{4585051 \,t^{10}}{572172412994582400}\,+
 \frac{5892151877\,t^{12}}{262340410524913476467712}\\
 \fl\qquad -\frac{586063249\,t^{14}}{10556078423502470838819840}+O\left(t^{16}\right),\\
 \fl \mathcal{B}_{_{\mathrm{sine}}}(5;t)=1-\frac{50\,t^2}{3267}+\frac{475\,t^4}{4008004}+O\left(t^6\right).
 \end{eqnarray*}

% \begin{thebibliography}{100}
  
 \end{document}